\documentclass[unnumsec,webpdf,modern,large]{oup-authoring-template}

\graphicspath{{Fig/}}

\usepackage[mathlines, switch]{lineno}
\usepackage{comment}
\usepackage[T1]{fontenc}

\usepackage{xurl}
\usepackage{hyperref}  

\theoremstyle{thmstyleone}%
\theoremstyle{thmstyletwo}%
\theoremstyle{thmstylethree}%

\begin{document}

\journaltitle{Preprint}
\copyrightyear{2026}
\pubyear{2026}

\firstpage{1}


\title[Molecular Fingerprints Are Strong Models for Peptide Function Prediction]{Molecular Fingerprints Are Strong Models for Peptide Function Prediction}

\author[$\ast,\dagger$]{Jakub Adamczyk\ORCID{0000-0003-4336-4288}}
\author[$\dagger$]{Piotr Ludynia\ORCID{0009-0004-0749-9569}}
\author[]{Wojciech Czech\ORCID{0000-0002-1903-8098}}

\authormark{Jakub Adamczyk et al.}

\address{\orgdiv{Faculty of Computer Science}, \orgname{AGH University of Krakow}, \orgaddress{\street{ al. Mickiewicza 30}, \postcode{30-059}, \state{Krakow}, \country{Poland}}}

\corresp[$\ast$]{Corresponding author, \href{email:jadamczy@agh.edu.pl}{jadamczy@agh.edu.pl}\\}
\corresp[$\dagger$]{Those authors contributed equally to this work.}


\received{Date}{0}{Year}
\revised{Date}{0}{Year}
\accepted{Date}{0}{Year}


\abstract{
\textbf{Motivation:} Understanding peptide properties is often assumed to require modeling long-range molecular interactions, motivating complex graph neural networks and pretrained transformers. Whether such long-range dependencies are essential remains unclear. We investigate if simple, domain-specific molecular fingerprints can capture peptide function without these assumptions. Atomic-level representations aim to provide richer information than purely sequence-based models and better efficiency than structural ones. \\
\textbf{Results:} Across 132 datasets, including LRGB and five additional peptide benchmarks, models using count-based ECFP, Topological Torsion, and RDKit fingerprints with LightGBM achieve state-of-the-art accuracy. Despite encoding only short-range molecular features, these models outperform GNNs and transformer-based approaches. Control experiments confirm that fingerprints, though inherently local, suffice for robust peptide property prediction. Our results challenge the presumed necessity of long-range interaction modeling and highlight molecular fingerprints as efficient, interpretable, and lightweight alternatives. \\
\textbf{Contact:} \href{jadamczy@agh.edu.pl}{jadamczy@agh.edu.pl}\\
\textbf{Supplementary information:} All code and data are available at: \\\url{https://github.com/scikit-fingerprints/peptides_molecular_fingerprints_classification}
}



\maketitle



\section{Introduction}

Peptides are short chains of amino acids, typically composed of 3–50 residues, that perform diverse biological functions and are increasingly being studied as potential therapeutics. They can bind to proteins, regulate their activity, and show antiviral, antitoxin, anticancer, and antidiabetic properties~\cite{peptides_drug_discovery, peptides_applications_diabetes}. In particular, antimicrobial peptides (AMPs) are promising candidates against the growing antimicrobial resistance crisis~\cite{peptides_applications_amp}. Therefore, accurate prediction of peptide properties is essential for drug discovery.

Peptide function prediction poses unique challenges for machine learning (ML), including homology bias and imbalanced datasets~\cite{AutoPeptideML,AMPBenchmark}. Consequently, numerous benchmark datasets with standardized evaluation procedures have been created. Peptides can be represented through various modalities, influencing the choice of ML models. When 3D structures are available, structural encodings can be used~\cite{PeptideReactor}, but these are rare and computationally expensive to obtain. Thus, sequence- and graph-based representations that avoid folding simulations have gained more attention.

Recent years have seen a shift from handcrafted sequence features to pretrained protein language models (PLMs) such as ProtBERT and ESM~\cite{ProtTrans,ESM2}. Alternatively, peptides can be modeled as molecular graphs, enabling the use of Graph Neural Networks (GNNs) and Graph Transformers (GTs). The Long-Range Graph Benchmark (LRGB)~\cite{LRGB} formulated peptide prediction as a long-range dependency problem. However, the practical importance of long-range interactions in short and flexible peptides remains unclear.

In chemoinformatics, molecular fingerprints - compact encodings derived from small molecular subgraphs \cite{molecular_descriptors} - are widely used for prediction of small-molecule properties. They offer domain-specific, efficient representations that often rival deep learning methods \cite{fingerprints_vs_gnns,molecular_property_predictions_gnn_limitations}. Despite their success, their use for peptide prediction has been limited, with only a few hybrid approaches that incorporate binary fingerprints \cite{MMDB,peptide_permeability_prediction}.

In this work, we revisit count-based molecular fingerprints as features for peptide function prediction. When combined with a LightGBM classifier, these representations achieve state-of-the-art results across six benchmarks, including LRGB, without requiring hyperparameter tuning or 3D structural information. This suggests that short-range subgraph statistics may be sufficient to capture key biochemical features, challenging the assumption that long-range dependencies are essential.

Our contributions are threefold, as we (1) demonstrate that count-based molecular fingerprints outperform alternative models on 132 datasets, (2) provide the most comprehensive benchmark of fingerprint-based peptide models to date, and (3) we analyze the proposed method in controlled experiments, showing its robustness.

\section{Related works}

Prediction approaches for peptides differ mainly by representation. For atom-level graphs, message-passing GNNs such as GCN, GraphSAGE, and GIN rely on local neighborhood aggregation \cite{GCN, GraphSAGE, GIN}. Graph Transformers replace message passing with attention mechanisms, typically with sparsified or semi-local variants, rather than fully global attention, to manage computational cost. Their rchitectures, like GraphGPS, HDSE, GRIT, GraphViT, and S$^{2}$GCN, aim to combine local and long-range modeling, and report strong performance on peptides datasets in LRGB benchmark \cite{rampavsek2022recipe, HDSE, GRIT, GraphViT, S2GCN}.

Sequence-based pipelines remain competitive, especially when data is scarce. They derive physicochemical and compositional descriptors (e.g. amphiphilicity, hydrophobicity) and use classical classifiers like RandomForest or SVM. Notable examples include Ampir, MACREL and amPEPpy \cite{ampir, MACREL, lawrence2021ampeppy}, and early deep learning models such as AMPScannerV2 \cite{veltri2018deep}. More recently Protein Language Models (PLMs), based on transformer architecture, such as ProtBERT and ProtT5 gained popularity, utilizing pretraining on large protein corpora \cite{ProtTrans}. The ESM family of models shows a scaling behavior similar to LLMs and give strong results on independent peptide benchmarks \cite{ESM2, AutoPeptideML}. For AMPs, specialized PLMs are available, e.g., BERT‑Protein, LMPred, cAMPs‑pred, and AMP‑BERT \cite{BERT_Protein, dee2022lmpred, ma2022identification, lee2023amp}.

Our work differs in adopting domain‑specific, count-based molecular fingerprints as peptide structure encoders. Compared with deep models, they provide robust, low‑parameter features and, as we show, give strong performance in peptide function prediction tasks without incorporating long‑range mechanisms.

\section{Methods}

Here, we describe the molecular fingerprints that form our proposed atom-level peptide encoding. We also describe the approach of Long Range Graph Benchmark to testing the presence of long-range dependencies.

\subsection{Molecular fingerprints}

\begin{figure}[b]
    \includegraphics[scale=0.4]{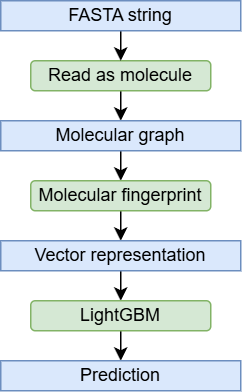}
    \centering
    \caption{The proposed pipeline for peptide function prediction.}
    \label{workflow_fig}
\end{figure}

Molecular fingerprints are feature extraction algorithms for molecules, based on extracting small subgraphs and detecting their presence (binary variant) or counting occurrences (count variant). This turns the problem of molecular graph classification into tabular classification, typically combined with tree-based ensembles such as Random Forest or LightGBM \cite{LightGBM}. See Figure \ref{workflow_fig} for visualization of the proposed pipeline.

We focus on hashed fingerprints \cite{molecular_descriptors}, exemplified by ECFP \cite{ECFP} or Topological Torsion \cite{topological_torsion}. They are more flexible than structural fingerprints like MACCS or PubChem, which use predefined subgraphs like functional groups, and are not designed for peptides. Hashed fingerprints use a more flexible approach, defining a general ``shape'' of extracted subgraphs, for example, circular atom neighborhoods in ECFP \cite{ECFP}, paths of given length in Topological Torsion (TT) \cite{topological_torsion}, or all small subgraphs in RDKit fingerprint \cite{rdkit_fp}. Extracting subgraph features in this manner avoids predefining features explicitly, offering greater flexibility. Subgraphs are defined by their structure, including the specific atoms and bonds they contain (e.g. characterized by atomic numbers and bond types), which are then converted into unique integer identifiers. To obtain a constant-length representation, they are hashed into the output vector, typically using the modulo function. The binary variant only indicates whether a given substructure appears in a molecule at all, while the count-based variant tallies the occurrences, incorporating more information about the compound composition and size.

\begin{figure}[b]
    \includegraphics[scale=0.45]{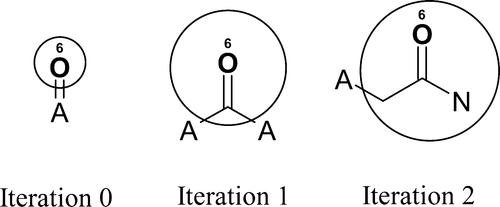}
    \centering
    \caption{Example subgraph extraction in consecutive iterations of ECFP algorithm. Reprinted with permission from \cite{ECFP}. Copyright 2010 American Chemical Society.}
    \label{fig:ecfp-iterations}
\end{figure}

ECFP typically uses circular subgraphs of radius 2 (see Figure \ref{fig:ecfp-iterations}). It starts with the atom itself, creating a numerical identifier based on atomic number, number of heavy neighbors, valence, and a few other simple features (see \cite{ECFP} for a complete list). Then, it iteratively increases the radius, incorporating neighbors by combining their identifiers, up to a given maximal radius. This results in the final subgraph identifier for each atom. Such a procedure, known as the Morgan algorithm, closely resembles the operation of message passing in GNNs, as both are rooted in the Weisfeiler-Lehman isomorphism test. Topological Torsion \cite{topological_torsion}, designed to model short-range molecular interactions, encodes linear paths of length 4, combining the identifiers of these consecutive atoms. RDKit fingerprint \cite{rdkit_fp} uses all subgraphs of size up to 7 bonds, which can be nonlinear and include cycles (e.g. rings). The major difference between TT and RDKit is that the former uses only linear paths, while the RDKit fingerprint also encodes small cyclic structures.

The shapes of subgraphs extracted by ECFP with radius 2 are roughly equivalent to those of a shallow 2-layer GNN \cite{GNNs_as_continous_fingerprints}. However, fingerprints do not learn task-specific weights like GNNs, instead performing domain-specific, deterministic feature extraction. This naturally reduces their risk of overfitting and increases the robustness of the learned feature space \cite{fingerprints_vs_gnns,molecular_property_predictions_gnn_limitations}, particularly beneficial in small, noisy datasets typical in chemo- and bioinformatics.

We focus on the three aforementioned fingerprints, as they are strictly local, very short-range descriptors. Therefore, a good performance of models based on their features can be attributed only to the importance of the short-range relations, rather than to long-range dependencies, since they are inherently incapable of capturing them. This locality makes them notably fast to compute \cite{scikit_fingerprints} and scalable to larger molecules like peptides, e.g. ECFP scales linearly with the number of atoms. Because peptides are relatively linear structures, path-based TT and RDKit fingerprints also scale well. Those three fingerprints have also shown remarkable performance in ligand-based virtual screening \cite{virtual_screening}.

\begin{figure}[b]
    \includegraphics[width=0.47\textwidth]{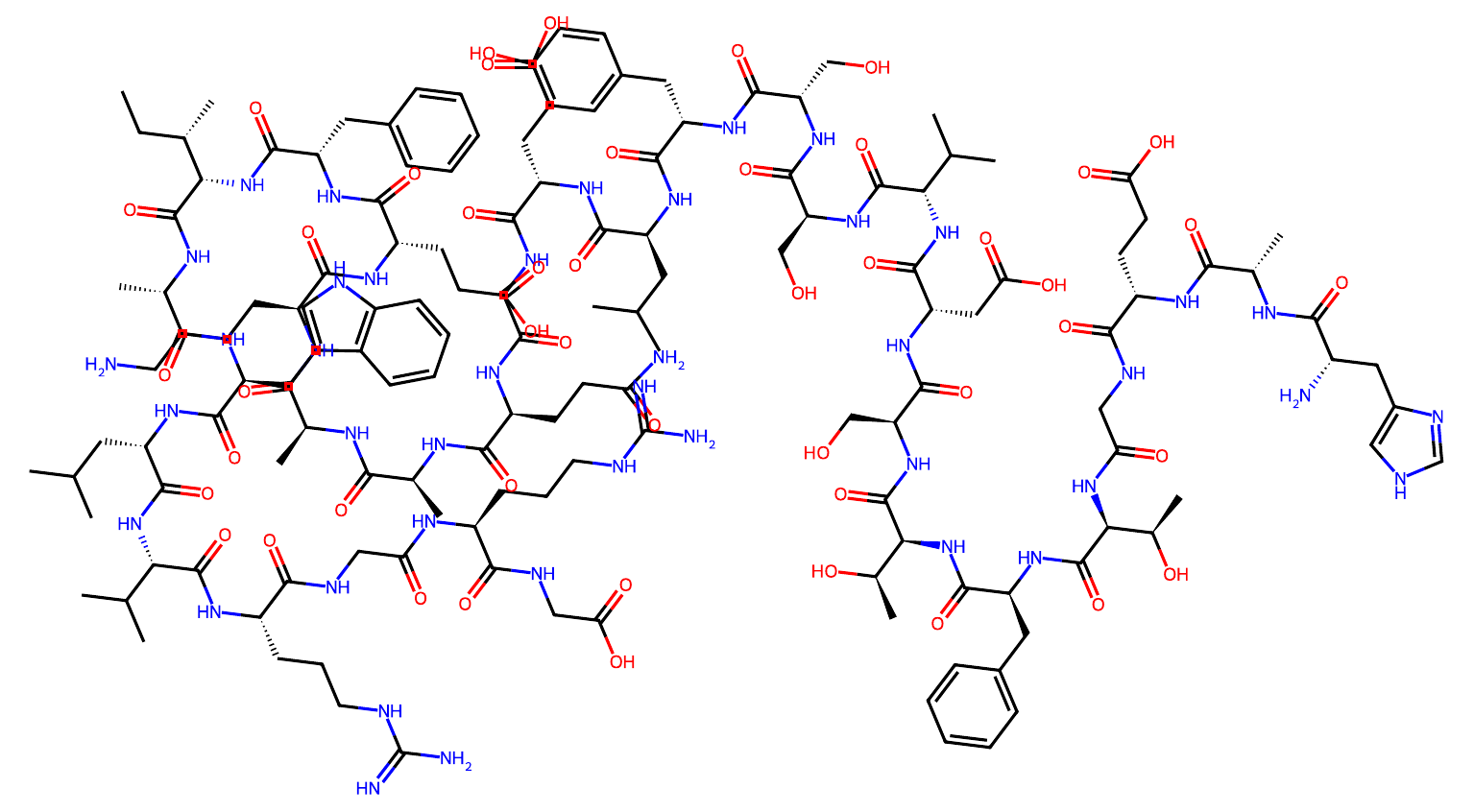}
    \centering
    \caption{Atom-level molecular graph for Liraglutide peptide.}
    \label{fig:peptide-graph}
\end{figure}

Lastly, we note that molecular fingerprints do not require any information about molecular conformation, peptide folding, or even whether a stable structure exists at all. Atom-level graphs (see Figure \ref{fig:peptide-graph} for an example) can be quickly and deterministically constructed from an amino acid sequence, and then vectorized by a fingerprint algorithm.

\subsection{Long-range interactions (LRI)}

Long Range Graph Benchmark (LRGB) \cite{LRGB} proposed 3 conditions for a dataset to potentially require learning long-range interactions (LRI). Firstly, a necessary condition is that the graphs must be sufficiently large, significantly exceeding the typical GNN receptive field. Secondly, the nature of the task itself must require long-range dependencies between nodes, with local interactions being insufficient. Lastly, the experiments should confirm this - models with global knowledge, such as GNNs with topological encodings or graph transformers, must outperform local message-passing ones. For graph classification and regression, peptide property prediction was proposed as a task family meeting these conditions.

We show that this is not necessarily the case, especially for the third condition. While atomic-level peptide graphs are large and elongated (large diameter, low atom degrees) enough to suggest potential LRI, the biological motivation for such interactions in peptide property prediction is weaker than for larger proteins. In proteins, LRI arise from complex folding driven by distant amino acid interactions and strongly influence functional properties. By contrast, the smaller size of peptides naturally limits these effects and often leads to a lack of well-defined structure \cite{peptides_structure}.

The third point, namely the superior performance of models with global knowledge, is the easiest to validate quantitatively. Since the argument is based on experimental results, it suffices to show that inherently local short-range models outperform long-range ones. Molecular fingerprints exactly fulfill these requirements. By obtaining SOTA results, we show that they provide strong evidence against the third claim, indicating that peptide function prediction primarily depends on short-range interactions.

\section{Experiments and results}

We train the proposed fingerprint-based models on a diverse set of benchmarks covering a wide range of peptide properties, data sampling strategies, evaluation metrics, and train-test splits. We closely follow the procedures of the original benchmark publications and report the corresponding metrics. When multiple metrics are available, we report AUROC and Matthews Correlation Coefficient (MCC) due to space constraints, with the remaining metrics provided in the Supplementary Material. The best result in each table is highlighted in bold.

Our proposed fingerprint-based models use count variants throughout and are implemented with the \texttt{scikit-fingerprints} \cite{scikit_fingerprints} library. We use default values for all hyperparameters: ECFP radius 2 (diameter 4, i.e., ECFP4), Topological Torsion path length 4, and RDKit fingerprint path length 7. They use small subgraphs as features, capturing only very short-range dependencies. We observed no significant performance gains from hyperparameter tuning (except for PeptideReactor, see Section \ref{section_general_peptide_benchmarks}).

We use LightGBM \cite{LightGBM} with 500 trees as the classifier due to its robust performance. Since most datasets are highly imbalanced, we apply class weighting inversely proportional to the positive class frequency. All other hyperparameters are kept at their default values, as tuning yielded negligible improvements. Increasing the number of trees beyond the default 100 was the only change that consistently improved performance, in line with prior literature \cite{tunability_probst}.

The proposed method's low sensitivity to hyperparameters is advantageous, as it reduces computational cost for new tasks, especially in large-scale virtual screening of peptide therapeutics. It also provides a fast and easy-to-use baseline for future works in this area.

\subsection{Long Range Graph Benchmark (LRGB)}

We first report results on the LRGB \cite{LRGB} benchmark for both datasets - \textit{Peptides-func} (classification) and \textit{Peptides-struct} (regression) - in Table \ref{table_results_lrgb}. These are multioutput datasets, and we report metrics averaged across all tasks. We compare against a broad set of graph-based architectures, including the original Graph Transformer and SAN \cite{LRGB}, as well as more recent models such as the MOLTOP baseline \cite{MOLTOP}, well-tuned classical GNNs (GCN, GatedGCN, GINE) \cite{LRGB_reassessment}, and models with enhanced long-range capabilities including GRIT \cite{GRIT}, HDSE \cite{HDSE}, and S$^2$GCN \cite{S2GCN}.

Since both datasets are multitask and LightGBM supports only single outputs, we train a separate model per task. For \textit{Peptides-struct}, we optimize MAE loss function. LRGB results for GNNs are reported over 10 random seeds. LightGBM with default settings is deterministic, so we do not report standard deviations. Results for Random Forest and Extremely Randomized Trees, which depend on random seed, are provided in the Supplementary Material. Their standard deviations are very low, $<0.1$ AUPRC.

The key observation is that all fingerprint-based models achieve SOTA results on both datasets. The strictly local ECFP model, based on radius 2 subgraphs, exceeds the best long-range GNN, S$^2$GCN, by $1.5\%$ AUPRC. Topological Torsion and RDKit fingerprint obtain similar results. The fact that these inherently short-range models reach such a performance contradicts the conclusions of \cite{LRGB} that incorporating long-range dependencies is necessary for accurate peptide function prediction. Instead, our results underscore the dominant role of short-range interactions.

\begin{table}[]
\centering
\caption{The results on LRGB benchmark.}
\begin{tabular}{|c|c|c|}
\hline
\textbf{Model} & \textbf{\begin{tabular}[c]{@{}c@{}}Peptides-func\\ AUPRC $\uparrow$\end{tabular}} & \textbf{\begin{tabular}[c]{@{}c@{}}Peptides-struct\\ MAE $\downarrow$\end{tabular}} \\ \hline
Transformer    & 63.26 $\pm$ 1.26                                                                  & 0.2529 $\pm$ 0.0016                                                                 \\ \hline
SAN            & 64.39 $\pm$ 0.75                                                                  & 0.2545 $\pm$ 0.0012                                                                 \\ \hline
MOLTOP         & 64.59 $\pm$ 0.05                                                                  & -                                                                                   \\ \hline
GraphGPS       & 65.35 $\pm$ 0.41                                                                  & 0.2500 $\pm$ 0.0005                                                                 \\ \hline
GINE           & 66.21 $\pm$ 0.67                                                                  & 0.2473 $\pm$ 0.0017                                                                 \\ \hline
GatedGCN       & 67.65 $\pm$ 0.47                                                                  & 0.2477 $\pm$ 0.0009                                                                 \\ \hline
GCN            & 68.60 $\pm$ 0.50                                                                  & 0.2460 $\pm$ 0.0007                                                                 \\ \hline
GraphViT       & 69.42 $\pm$ 0.75                                                                  & 0.2449 $\pm$ 0.0016                                                                 \\ \hline
GRIT           & 69.88 $\pm$ 0.82                                                                  & 0.2460 $\pm$ 0.0012                                                                 \\ \hline
CRaWl          & 70.74 $\pm$ 0.32                                                                  & 0.2506 $\pm$ 0.0022                                                                 \\ \hline
GRED           & 71.33 $\pm$ 0.11                                                                  & 0.2455 $\pm$ 0.0013                                                                 \\ \hline
DRew           & 71.50 $\pm$ 0.44                                                                  & 0.2536 $\pm$ 0.0015                                                                 \\ \hline
HDSE           & 71.56 $\pm$ 0.58                                                                  & 0.2457 $\pm$ 0.0013                                                                 \\ \hline
S$^2$GCN       & 73.11 $\pm$ 0.66                                                                  & 0.2447 $\pm$ 0.0032                                                                 \\ \hline
\hline
RDKit          & 73.11                                                                  & 0.2459                                                                 \\ \hline
TT             & 73.18                                                                  & 0.2438                                                                 \\ \hline
\textbf{ECFP}  & \textbf{74.60}                                                         & \textbf{0.2432}                                                        \\ \hline
\end{tabular}
\label{table_results_lrgb}
\end{table}

\begin{table}[]
\caption{Differences between binary and count fingerprint models.}
\centering
\begin{tabular}{|cc|c|c|c|}
\hline
\multicolumn{1}{|c|}{\textbf{Dataset}} & \textbf{Variant} & \textbf{ECFP} & \textbf{TT} & \textbf{RDKit} \\ \hline
\multicolumn{1}{|c|}{Peptides-func}    & binary           & 70.57         & 66.18       & 63.88          \\ \hline
\multicolumn{1}{|c|}{Peptides-func}    & binary + length          & 68.46         & 66.49       & 66.97          \\ \hline
\multicolumn{1}{|c|}{Peptides-func}    & count            & 74.60         & 73.18       & 73.11          \\ \hline
\multicolumn{2}{|c|}{AUPRC gain $\uparrow$}               & +4.03         & +7.00       & +9.23          \\ \hline
\hline
\multicolumn{1}{|c|}{Peptides-struct}  & binary           & 0.3049        & 0.3298      & 0.3331         \\ \hline
\multicolumn{1}{|c|}{Peptides-struct}  & binary + length           & 0.2652        & 0.2691      & 0.2744         \\ \hline
\multicolumn{1}{|c|}{Peptides-struct}  & count            & 0.2432        & 0.2438      & 0.2459         \\ \hline
\multicolumn{2}{|c|}{MAE gain $\downarrow$}               & -0.0617       & -0.0860     & -0.0872        \\ \hline
\end{tabular}
\label{table_count_vs_binary}
\end{table}

\begin{table*}[]
\centering
\caption{The results on BERT-based models benchmark \cite{BERT_benchmark}.}
\begin{tabular}{|c|c|c|c|c|c|c|c|}
\hline
\textbf{Model} & \textbf{ADAPTABLE} & \textbf{APD}  & \textbf{CAMP} & \textbf{dbAMP} & \textbf{DRAMP} & \textbf{YADAMP} & \textbf{Average F1} \\ \hline
AMP-BERT & 81.7 & 78.3 & 82.3 & 62.7 & 64.0 & 82.6 & 75.3 \\ \hline
BERT-Protein & 79.4 & 90.1 & 80.6 & 82.9 & 58.8 & 58.8 & 75.1 \\ \hline
cAMPs\_pred & 61.3 & 74.1 & 72.8 & 52.4 & 39.2 & 85.0 & 64.1 \\ \hline
LM\_pred & 55.9 & 71.7 & 64.5 & 60.7 & 47.6 & 82.1 & 63.8 \\ \hline
LM\_pred (BFD) & 73.9 & 89.6 & 83.6 & 68.7 & 64.8 & 84.4 & 70.3 \\ \hline
\hline
RDKit & \textbf{82.0} & 93.0 & 88.4 & 91.4 & 76.4 & 92.0 & 87.2 \\ \hline
TT & 80.6 & \textbf{94.3} & \textbf{88.8} & 90.8 & \textbf{76.9} & \textbf{95.8} & \textbf{87.8} \\ \hline
ECFP & 80.8 & 93.0 & 88.2 & \textbf{91.5} & 76.3 & \textbf{95.8} & 87.6 \\ \hline
\end{tabular}
\label{table_results_bert_benchmark}
\end{table*}

The ECFP4 fingerprint functions similarly to a two-layer message-passing GNN, yet it outperforms both local models (e.g., GCN) and long-range variants such as S$^2$GCN and HDSE. We attribute this to three factors.

First, fingerprints count discrete substructures rather than learning continuous graph representations. This provides a stronger inductive bias and reduces reliance on large training datasets.

Second, peptides consist of repeating fragments, amino acids, that are structurally similar and recur frequently. Thus, models mainly need to detect recurring small subgraphs (e.g., amine groups, \texttt{-NH2}). This is not indicative of genuine long-range interaction, but rather reflects the straightforward repetition of common motifs across different regions of the input.

Third, count-based fingerprints naturally capture effects of structure size and substructure frequency. In \textit{Peptides-struct}, regression targets such as inertia mass and length strongly correlate with peptide size. In \textit{Peptides-func}, classification labels (e.g., antimicrobial activity) depend on binding properties linked to surface area and thus size. These properties are therefore more closely tied to local structures and overall peptide size than to long-range interactions, and can be well approximated by counting subgraphs.

Our results therefore indicate that the \textit{Peptides-func} and \textit{Peptides-struct} datasets do not require modeling long-range interactions. Previous studies may have overlooked this, as \cite{LRGB} evaluated only GNNs, and \cite{LRGB_reassessment} focused on deep models (6 to 10 layers) with global topological encodings. This underscores the importance of comparing against established, domain-specific baselines for a fair evaluation. To emphasize the effect of count-based fingerprints, we compare them with (1) binary variants (2) binary variants with sequence length feature, see Table \ref{table_count_vs_binary}. We observe substantial performance gains from using count features instead of binary ones. They also strongly outperform binary fingerprints with sequence length, meaning that the count-based model does not simply approximate the overall peptide size. Given those findings, it is concerning that all hybrid models in the bioinformatics literature we reviewed \cite{MMDB,peptide_permeability_prediction} rely exclusively on binary fingerprints, despite count variants having identical computational cost and markedly better performance.

We also highlight the efficiency of our approach. In terms of wall-clock time, computing ECFP fingerprints and training LightGBM takes 19 seconds on a 12-core Intel Core i7-12700KF CPU for \textit{Peptides-func}. In contrast, training the SAN graph transformer requires up to 60 hours on an NVIDIA A100 GPU \cite{LRGB}. Additional timing results are provided in the Supplementary Material.

\subsection{Antimicrobial peptides (AMPs) benchmarks}

In subsequent experiments, we examine whether the LRGB results generalize to other benchmarks in this domain. Such generalization would indicate that peptide function prediction is not inherently dependent on long-range interactions.

We first report results on antimicrobial peptide (AMP) prediction benchmarks. Although those benchmarks target the same property, antimicrobial activity, they differ in structural diversity, construction of positive and negative examples, and train-test splitting procedures. We consider three benchmarks that span these variations: the BERT-based models benchmark \cite{BERT_benchmark}, XUAMP \cite{Xu_AMP}, and AMPBenchmark \cite{AMPBenchmark}. Notably, many datasets in this section include peptides that exceed the typical length definition, with averages of 70-100 amino acids. Achieving strong performance therefore requires fingerprint-based models to be both scalable, due to atom-level processing, and effective on substantially larger molecules. We include detailed dataset descriptions and sequence length statistics in Supplementary Material.

\subsubsection{BERT-based models benchmark}

First, we compare our approach with BERT-based models pretrained for AMP prediction, as evaluated in \cite{BERT_benchmark}. This benchmark includes AMPs from six classic datasets, uses negative samples from UniProt, and applies CD-HIT \cite{CDHIT} with a $40\%$ similarity threshold to reduce homology bias. No additional fine-tuning is performed, as the BERT models are already trained specifically for AMP prediction, including classification heads. The authors also note that parts of these datasets may be present in the pretraining data, which likely benefits the BERT models on this benchmark. To mimic such pretraining with fingerprints and LightGBM, we adopt a leave-one-dataset-out strategy. For example, when evaluating on the ADAPTABLE dataset, we merge all remaining datasets into a single training set and apply CD-HIT-2D \cite{CDHIT} at a $40\%$ threshold to remove peptides overly similar to those in the test set.

Results are reported in Table \ref{table_results_bert_benchmark}, focusing on F1 score, as all models achieve nearly identical AUROC values close to $100\%$. Additional metrics are provided in the Supplementary Material. Fingerprint-based models achieve state-of-the-art performance on all datasets, with Topological Torsion (TT) yielding the best average F1 score. Performance differences vary by dataset but show no clear dependence on test set size or peptide size. For instance, ADAPTABLE and DRAMP both contain approximately 4000 peptides with an average length of $75$ amino acids, yet differ substantially in difficulty. On ADAPTABLE, fingerprints improve F1 by $0.3\%$ over the best BERT-based model, while on DRAMP the gain reaches $12.1\%$. These results demonstrate that atom-level, short-range fingerprints consistently yield strong models, but the exact data used for training has a considerable impact on the final performance.

\subsubsection{XUAMP benchmark}

Datasets used in the previous section are widely adopted in AMP research, but are relatively small. To address this limitation, a unified AMP benchmarking dataset known as XUAMP was proposed \cite{Xu_AMP}. It is constructed from nine commonly used datasets, such as CAMP and DRAMP, which are merged and deduplicated using homology-based CD-HIT clustering \cite{CDHIT}. This prevents data leakage by ensuring that highly similar sequences do not appear in both training and test sets, ensuring a realistic evaluation. The authors benchmarked traditional feature engineering approaches combined with various classifiers, including Random Forest, SVM, and k nearest neighbors (kNN).

We exactly replicate the original XUAMP setup and data splits, and report the results in Table \ref{table_results_xu_amp}. Fingerprint-based models again achieve SOTA performance, demonstrating robustness across different dataset sizes. The strongest prior method, AMPfun \cite{AMPfun}, relies on a relatively complex pipeline that combines binary amino acid position profiles, sequence composition features, physicochemical descriptors, and feature selection. Such pipelines can be fragile, as evidenced by the superior performance of the much simpler ECFP fingerprint in both AUROC and MCC.

\begin{table}[]
\centering
\caption{The results on XUAMP benchmark \cite{Xu_AMP}.}
\begin{tabular}{|c|c|r|}
\hline
\textbf{Method} & \textbf{AUROC} & \multicolumn{1}{c|}{\textbf{MCC}}   \\ \hline
AMPscannerV2    & 58.5           & 0.137                               \\ \hline
iAMP-2L         & 59.2           & 0.261                               \\ \hline
ADAM-SVM        & 61.2           & 0.264                               \\ \hline
ampir           & 61.9           & 0.156                               \\ \hline
MLAMP           & 62.9           & 0.194                               \\ \hline
ADAM-HMM        & 68.4           & 0.39                                \\ \hline
AMPlify         & 69.7           & 0.381                               \\ \hline
AMPEP           & 72.7           & 0.425                               \\ \hline
AMPfun          & 73.5           & 0.414                               \\ \hline
\hline
RDKit           & 74.5           & \multicolumn{1}{c|}{0.404}          \\ \hline
TT              & 73.1           & \multicolumn{1}{c|}{0.385}          \\ \hline
\textbf{ECFP}            & \textbf{75.0}  & \multicolumn{1}{c|}{\textbf{0.429}} \\ \hline
\end{tabular}
\label{table_results_xu_amp}
\end{table}

\subsubsection{AMPBenchmark}

The authors of AMPBenchmark \cite{AMPBenchmark} observed that negative class sampling has a strong impact on model performance. Their benchmark evaluates the sensitivity of AMP predictors to data choice by retraining models across a large collection of datasets with varying negative sampling strategies. Evaluated  models range from sequence-based feature engineering approaches, such as AmpGram \cite{AmpGram}, through CNN-based deep learning models like Deep-AmPEP30 \cite{DeepAmPEP30}, to multimodal ensembles combining physicochemical and sequence features, such as ampir \cite{ampir}.

Model robustness is assessed by retraining on a suite of datasets constructed by pairing the same positive class peptides with 11 different negative sampling strategies, each repeated five times using distinct UniProt subsets. This yields 55 datasets that vary in sequence length distributions, sequence similarity, and class balance, each evaluated using an 80-20\% train-test split. Results on AMPBenchmark are summarized in Table \ref{table_results_ampbenchmark_sidorczuk}, reporting the mean and standard deviation across all retraining runs. Once again, fingerprint-based models achieve the best performance, with substantially lower standard deviations than competing methods. This suggests that feature engineering based on peptide topological graph yields more robust and reliable models than complex ensembles. Despite their popularity in bioinformatics, the latter are inherently susceptible to data sampling strategy and human bias in descriptor selection.

\begin{table}[]
\centering
\caption{The results on AMPBenchmark \cite{AMPBenchmark}.}
\begin{tabular}{|c|c|}
\hline
\textbf{Model} & \textbf{AUROC}       \\ \hline
AmPEP        & 61.69 $\pm$ 6.93          \\ \hline
iAMP-2L      & 62.60 $\pm$ 9.51           \\ \hline
SVM-LZ       & 79.71 $\pm$ 2.43          \\ \hline
CS-AMPPred   & 86.69 $\pm$ 4.36          \\ \hline
AMAP         & 89.87 $\pm$ 4.18          \\ \hline
Deep-AmPEP30 & 90.98 $\pm$ 5.61          \\ \hline
AmPEPpy      & 95.35 $\pm$ 2.99          \\ \hline
AmpGram      & 95.56 $\pm$ 2.54          \\ \hline
MACREL       & 96.29 $\pm$ 2.43          \\ \hline
MLAMP        & 96.68 $\pm$ 2.35          \\ \hline
ampir        & 96.71 $\pm$ 2.08          \\ \hline
AMPScannerV2 & 96.79 $\pm$ 2.08          \\ \hline
\hline
TT           & 97.19 $\pm$ 1.89          \\ \hline
RDKit        & 97.25 $\pm$ 1.84          \\ \hline
\textbf{ECFP}         & \textbf{97.37 $\pm$ 1.74} \\ \hline
\end{tabular}
\label{table_results_ampbenchmark_sidorczuk}
\end{table}

\subsection{General peptide benchmarks}
\label{section_general_peptide_benchmarks}

Beyond antimicrobial activity, many other peptide properties are of interest, including antioxidant activity, anti-MRSA effects, toxicity, and more. A common challenge in this area is the small size of available datasets, which often contain fewer than 1000 positive samples. To assess how fingerprint-based models perform under these constraints, we selected two large benchmarks comprising a total of 68 datasets that span a wide range of peptide properties: AutoPeptideML \cite{AutoPeptideML} and PeptideReactor \cite{PeptideReactor}.

\subsubsection{AutoPeptideML}

AutoPeptideML \cite{AutoPeptideML} comprises 18 datasets. Positive peptides are drawn from existing literature datasets, while negative ones are carefully filtered to avoid false negatives and sampled to match the sequence length distribution of the positives. Train-test splits are created using homology clustering to produce an out-of-distribution test set, analogous to scaffold splits commonly used in molecular property prediction. Together, these choices result in a particularly challenging evaluation setup. AutoPeptideML reports results for state-of-the-art protein language models (PLMs), including models with up to 3B parameters, such as Prot-T5-XL \cite{ProtTrans}. Their embeddings are combined with a complex ensemble of 30 classifiers: 10 k nearest neighbors (kNN), 10 Random Forests, and 10 LightGBMs, each tuned separately via Bayesian hyperparameter optimization.

Results are summarized in Table \ref{table_results_autopeptideml}. Fingerprint-based models are competitive in both MCC and AUROC with the best performing PLMs. Importantly, they are far more parameter-efficient, reaching comparable performance with roughly 22k parameters (counting tree nodes in LightGBM), compared to PLMs with up to 3 billion weights. Notably, the ECFP-based classifier outperforms three variants of ESM2 \cite{ESM2} in both AUROC and MCC, and achieves MCC nearly identical to ESM2-650M. A likely explanation is that peptides, unlike most proteins used to train and evaluate PLMs, are very short sequences. Fingerprints operate on full molecular graphs and can be viewed as a higher-resolution representation of peptides than sequence-only models.

\begin{table}[]
\centering
\caption{The results on AutoPeptideML benchmark \cite{AutoPeptideML}.}
\begin{tabular}{|c|c|c|c|}
\hline
\textbf{Model} & \textbf{\# params} & \textbf{AUROC} & \textbf{MCC}   \\ \hline
ProtBERT       & 420M               & 75.9           & 0.375          \\ \hline
ESM2-150M      & 150M               & 77.7           & 0.402          \\ \hline
Prost-T5       & 3B                 & 77.1           & 0.409          \\ \hline
ESM2-8M        & 8M                 & 77.5           & 0.418          \\ \hline
ESM2-35M       & 35M                & 78.0           & 0.428          \\ \hline
ESM1b-650M     & 650M               & 78.9           & 0.433          \\ \hline
ESM2-650M      & 650M               & \textbf{79.7}  & 0.438          \\ \hline
Prot-T5-XL     & 3B                 & \textbf{79.7}  & \textbf{0.447} \\ \hline
\hline
RDKit          & \textbf{20k}       & 76.9           & 0.421          \\ \hline
TT             & \textbf{23k}       & 77.1           & 0.422          \\ \hline
ECFP           & \textbf{22k}       & 78.1           & 0.437          \\ \hline
\end{tabular}
\label{table_results_autopeptideml}
\end{table}

\subsubsection{PeptideReactor}
\label{section_PeptideReactor}

PeptideReactor \cite{PeptideReactor} is a large benchmark comprising 50 datasets designed to compare peptide encoding methods. To focus on the impact of feature engineering, all models use the same Random Forest classifier with 100 trees. No classifier hyperparameter tuning or class weighting is applied, and only the feature extraction methods are tuned. The benchmark includes 48 encodings based on sequence information or structural descriptors derived from 3D structures. This setting enables a direct comparison of molecular fingerprints with a broad range of established peptide encodings from the bioinformatics literature.

For this benchmark, we perform some limited hyperparameter tuning for fingerprints. We make this exception, since authors explicitly encourage tuning for feature extraction, and perform it for many other methods, so tuning fingerprints results in more fair comparison. Further, many datasets in this benchmark contain relatively large proteins, which may benefit from slightly larger subgraphs in fingerprints. Specifically, 17 of the 50 datasets have average sequence lengths exceeding 50 amino acids, the typical upper limit for peptides, and 8 exceed 200 amino acids. See Supplementary Material for detailed statistics of all datasets. Using 5-fold cross-validation, we tune the ECFP radius in the range $[2, 4]$, the Topological Torsion path length in the range $[4, 6]$, and the RDKit path length in the range $[7, 9]$. These remain short-range descriptors, but allow slightly larger substructures to be captured.

Since this is an encoding benchmark, we also evaluate a general ``FP encoding'' setting, where the fingerprint type itself is treated as a hyperparameter. This allows us to address the question \textit{``How effective are encodings based on the topological graph (2D structure) of a molecule?''}. Such comparison is relevant because existing feature groups rely either on amino acid sequences (1D encodings) or spatial peptide structures (3D encodings). It is also a highly realistic setting, as the best encoding method is selected for each dataset. Results are summarized in Table \ref{table_results_peptidereactor}. Due to space constraints, we report only the five best sequence- and structure-based encodings here, with full results provided in the Supplementary Material.

Fingerprints significantly outperform all structure-based encodings that rely on spatial information, despite modeling only topological molecular graphs and short-range interactions. This suggests that these smaller proteins may exhibit biological behavior that differs from that of large proteins, for which spatial folding is known to be critical. Moreover, constructing molecular graphs takes only a few seconds, whereas spatial encodings require costly tertiary structure estimation, taking approximately 85 minutes per dataset \cite{PeptideReactor}.

Fingerprints are also highly competitive with the strongest sequence-based methods. Notably, the ECFP fingerprint alone ranks fourth overall in the benchmark, outperforming 45 other encodings. We also observe substantial variation in fingerprint performance across datasets. When treating the fingerprint type as a hyperparameter, the FP encoding achieves state-of-the-art performance. This indicates that fingerprints provide robust peptide representations that scale well across diverse properties and protein sizes, but the optimal choice depends on the data and task. We note that \cite{PeptideReactor} concludes that structural encodings are inferior to sequence-based ones in both computational cost and performance. Fingerprint-based encodings offer a counterexample to this conclusion, albeit relying on a topological 2D structure rather than conformational (spatial) information.

\begin{table}[]
\centering
\caption{The results on PeptideReactor benchmark \cite{PeptideReactor}.}
\begin{tabular}{|c|c|c|}
\hline
\textbf{Model group}         & \textbf{Model} & \textbf{Average F1} \\ \hline
\multirow{5}{*}{Structure}   & disord         & 53.5                \\ \cline{2-3} 
                             & electr         & 60.9                \\ \cline{2-3} 
                             & distan         & 61.3                \\ \cline{2-3} 
                             & delaun         & 69.5                \\ \cline{2-3} 
                             & qsar           & 72.8                \\ \hline
\multirow{5}{*}{Sequence}    & dde            & 81.3                \\ \cline{2-3} 
                             & ngram\_        & 81.6                \\ \cline{2-3} 
                             & psekraac       & 81.8                \\ \cline{2-3} 
                             & dist\_f        & 82.2                \\ \cline{2-3} 
                             & cksaap         & 82.4                \\ \hline
\multirow{4}{*}{Fingerprint} & RDKit          & 80.6                \\ \cline{2-3} 
                             & TT             & 80.6                \\ \cline{2-3} 
                             & ECFP           & 81.7                \\ \cline{2-3} 
                             & \textbf{FP encoding}    & \textbf{82.9}       \\ \hline
\end{tabular}
\label{table_results_peptidereactor}
\end{table}

\section{Additional baselines}

Two commonly used sequence-based baselines for proteins and peptides are amino acid counts and ESM2 \cite{ESM2}. The former uses a bag-of-words on sequences, typically including uni- and bigram counts. ESM2 embeds a peptide by extracting amino acid embeddings and combining them with mean pooling.

Note that molecular fingerprints can be viewed as a higher-resolution alternative to those methods. They operate directly on atomic graph, rather than on sequences, which enriches the data representation, particularly in realistic low-data regimes. This structural encoding incorporates information about amino acids, such as their elemental composition and charge, capturing structural similarities between amino acids. For example, they encode the fact that asparagine and glutamine are more similar to each other than asparagine and proline. In contrast, sequence-level models must learn these relationships implicitly, often requiring more data.

We compare the two baselines to ECFP fingerprints, with results shown in Table \ref{table_results_baselines}. LightGBM classifier is used in all cases except PeptideReactor, where we apply Random Forest as before. We follow the same data splits and evaluation metrics as in the earlier sections. We use the ESM2-35M variant, as larger models did not improve performance, consistent with prior findings \cite{AutoPeptideML}.

Molecular fingerprints give the best results in 5 out of 7 cases. They are particularly strong on PeptideReactor, with $4.9\%$ F1-score gain over ESM2. Further, they are always better than amino acid counts, and as such form a strong baseline for peptide property prediction.

\begin{table}[]
\caption{Results of additional baselines.}
\resizebox{0.48\textwidth}{!}{
\begin{tabular}{|c|c|c|c|c|}
\hline
\textbf{Benchmark} & \textbf{Metric} & \textbf{\begin{tabular}[c]{@{}c@{}}Amino acid\\ counts\end{tabular}} & \textbf{ESM2} & \textbf{ECFP} \\ \hline
AMPBenchmark & AUROC & 96.9\% & 97.3\% & \textbf{97.4\%} \\ \hline
AutoPeptideML & AUROC & 74.6\% & 77.2\% & \textbf{78.1\%} \\ \hline
BERT benchmark & F1 & 87.5\% & \textbf{89.8\%} & 87.6\% \\ \hline
LRGB, Peptides-func & AUPRC & 69.8\% & 66.0\% & \textbf{74.6\%} \\ \hline
LRGB, Peptides-struct & MAE & 0.2665 & 0.2828 & \textbf{0.2432} \\ \hline
PeptideReactor & F1 & 80.0\% & 76.8\% & \textbf{81.7\%} \\ \hline
XUAMP & AUROC & 74.2\% & \textbf{76.8\%} & 75.0\% \\ \hline
\end{tabular}
}
\label{table_results_baselines}
\end{table}

\section{Sequence shuffling}

To verify that short-range dependencies are sufficient, we conduct sequence shuffling experiments. A given percentage of amino acids in each training sequence is randomly permuted, acting as an adversarial perturbation. As this shuffling ratio increases, long-range interactions are progressively destroyed. If performance remains high even with fully randomized sequences, this indicates that peptide properties mainly depend on substructural composition. Additional experiments (shuffling performance for baselines, shuffling the test sequences) and detailed tables are reported in the Supplementary Material.

\begin{figure}[t]
    \centering
    \includegraphics[width=0.48\textwidth]{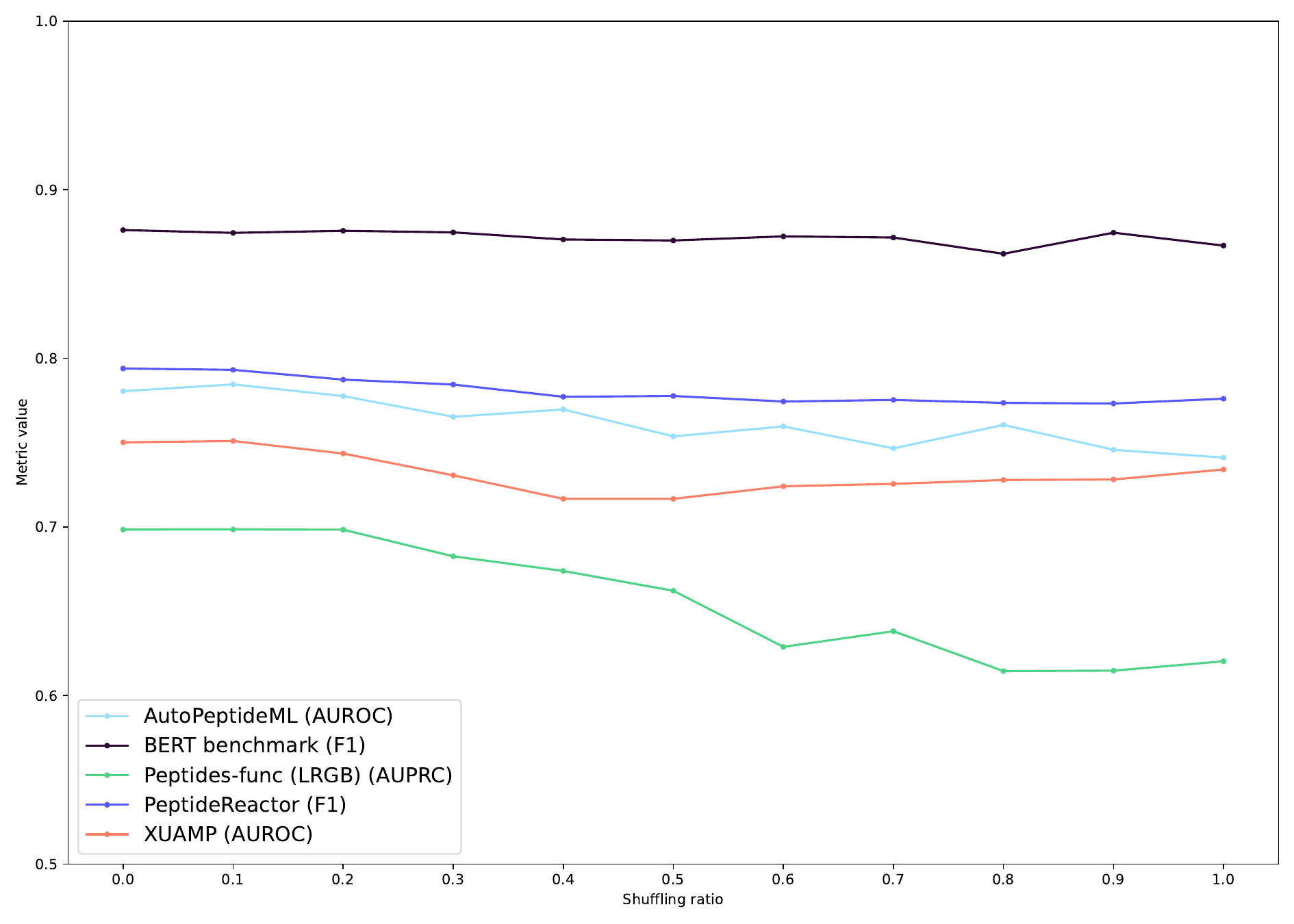}
    \caption{ECFP performance under increasing shuffling ratio.}
    \label{fig:shuffling-metrics}
\end{figure}

We use the ECFP fingerprint with default hyperparameters and report the average metric across each benchmark\footnote{We exclude AMPBenchmark due to its very large size.} for every shuffling ratio in Figure \ref{fig:shuffling-metrics}. Performance degradation for fingerprint-based models is small, even with fully randomized sequences, and does not exceed about $4\%$. On XUAMP, results are nearly identical to those obtained without shuffling. Thus, even when sequences are completely randomized and only very short-range interactions remain, the ECFP fingerprint maintains strong performance. This further supports the hypothesis that peptide property prediction primarily depends on local, atom-level subgraphs and also demonstrates the robustness of the proposed approach.

\section{Additional experiments}

Due to word limitations, we relegate a series of additional experiments to the Supplementary Materials. They further explore the robustness and possible limitations of the proposed approach. Here, we provide summary results.

First, we analyze error distributions and correlate them to physicochemical properties of peptides, but results do not indicate any particular patterns. Further, we explore binning data by length, e.g., testing only on shortest or longest peptides. All models show similar performance patterns, e.g., fingerprints and ESM2 underperform on shortest peptides, which have the least structural information.

Lastly, we purposefully design a long-range task, in order to delineate the limitations of the proposed method. The task is recognizing highly charged sequence motifs like ``KKK'' or ``RRR'' in sequence, which are known to impact peptide-protein binding and are highly relevant to peptide therapeutic design. It is trivial for sequence-based models like ESM2, which achieve near-perfect performance. However, it is very challenging for molecular fingerprints, as it requires using long-range and order-dependent on molecular graphs.


\section{Conclusions}

We presented an approach to peptide function prediction based on count variants of hashed molecular fingerprints. These methods use atom-level representations derived from molecular graphs, which are inexpensive to construct, fast to compute, and do not require knowledge of the folded peptide structure. Despite relying strictly on local, short-range descriptors, they achieve strong performance across a broad range of tasks.

When combined with LightGBM, this approach delivers superior results in a large-scale evaluation spanning six benchmarks and 132 datasets. They outperform a wide range of alternatives, including GNNs, pretrained amino acid sequence transformers, hand-crafted feature engineering pipelines, and multimodal models. Beyond predictive quality, the proposed models are computationally lightweight.

To the best of our knowledge, this work provides the most extensive comparison of molecular fingerprints applied to peptides to date. In contrast to prior bioinformatics studies, we focus on count-based rather than binary fingerprints and show that this design choice is critical for strong performance. Our main methodological contribution is demonstrating that inherently short-range models achieve state-of-the-art results on peptide property prediction tasks. In particular, an ECFP-based classifier with radius 2 surpasses the previous best model on \textit{Peptides-func} dataset from LRGB benchmark by $1.5\%$ AUPRC. These findings challenge the presumed necessity of long-range dependencies for this task and reinforce previous observations in the literature \cite{LRGB_reassessment}.

Future work will extend this approach to property prediction for larger proteins. We also plan to analyze fingerprint performance on tasks involving chemically modified peptides, such as cyclic peptides, which are of high interest for drug design and require atom-level featurization.

In conclusion, atom-level, short-range models based on molecular fingerprints constitute strong and reliable approaches for peptide property prediction. Their use is also essential for critically evaluating claims about the importance of long-range dependencies in molecular learning tasks.


\section{Acknowledgements}

The authors thank the BIT Student Scientific Group for computational resources and assistance 
for this project. The authors thank Aleksandra Elbakyan for her work and support for accessibility of science.

\subsection{Data availability}

All datasets and benchmarks used are publicly available at \\ \url{https://github.com/scikit-fingerprints/peptides_molecular_fingerprints_classification}.

\section{Author contributions statement}

Contributions follow the CRediT format. J.A. was responsible for conceptualization, methodology, software, validation, investigation, data curation, and writing. P.L. was responsible for software, investigation, and writing. W.C. was responsible for writing, review, and supervision.

\section{Competing interests}

No competing interest is declared.

\section{Funding}

This work was supported by the general funding from the Ministry of Science and Higher Education to AGH University in Krakow, and by project Excellence Initiative – Research University (IDUB) for AGH University of Krakow.

\bibliographystyle{plain}
\bibliography{bibliography}

@article{MMDB,
title = {MMDB: Multimodal dual-branch model for multi-functional bioactive peptide prediction},
journal = {Analytical Biochemistry},
volume = {690},
pages = {115491},
year = {2024},
issn = {0003-2697},
author = {Yan Kang and Huadong Zhang and Xinchao Wang and Yun Yang and Qi Jia},
}

@article{epps_singleton_test,
    title={An omnibus test for the two-sample problem using the empirical characteristic function},
    author={Epps, Thomas W and Singleton, Kenneth J},
    journal={Journal of Statistical Computation and Simulation},
    volume={26},
    number={3-4},
    pages={177-203},
    year={1986},
    publisher={Taylor \& Francis}
}

@article{peptide_permeability_prediction,
author={Tan, Xiaorong
and Liu, Qianhui
and Fang, Yanpeng
and Zhu, Yingli
and Chen, Fei
and Zeng, Wenbin
and Ouyang, Defang
and Dong, Jie},
title={Predicting Peptide Permeability Across Diverse Barriers: A Systematic Investigation},
journal={Molecular Pharmaceutics},
year={2024},
month={Aug},
day={05},
publisher={American Chemical Society},
volume={21},
number={8},
pages={4116-4127},
issn={1543-8384},
}

@article{molecular_property_predictions_gnn_limitations,
  title={{Understanding the Limitations of Deep Models for Molecular property prediction: Insights and Solutions}},
  author={Xia, Jun and Zhang, Lecheng and Zhu, Xiao and Liu, Yue and Gao, Zhangyang and Hu, Bozhen and Tan, Cheng and Zheng, Jiangbin and Li, Siyuan and Li, Stan Z},
  journal={Advances in Neural Information Processing Systems},
  volume={36},
  pages={64774--64792},
  year={2023}
}

@article{peptides_applications_diabetes,
author={Wang, Lei
and Wang, Nanxi
and Zhang, Wenping
and Cheng, Xurui
and Yan, Zhibin
and Shao, Gang
and Wang, Xi
and Wang, Rui
and Fu, Caiyun},
title={Therapeutic peptides: current applications and future directions},
journal={Signal Transduction and Targeted Therapy},
year={2022},
month={Feb},
day={14},
volume={7},
number={1},
pages={48},
issn={2059-3635},
}

@article{peptides_applications_amp,
author={Szymczak, Paulina
and Mo{\.{z}}ejko, Marcin
and Grzegorzek, Tomasz
and Jurczak, Rados{\l}aw
and Bauer, Marta
and Neubauer, Damian
and Sikora, Karol
and Michalski, Micha{\l}
and Sroka, Jacek
and Setny, Piotr
and Kamysz, Wojciech
and Szczurek, Ewa},
title={Discovering highly potent antimicrobial peptides with deep generative model HydrAMP},
journal={Nature Communications},
year={2023},
month={Mar},
day={15},
volume={14},
number={1},
pages={1453},
issn={2041-1723},
}

@article{peptides_drug_discovery,
    author = {Goles, Montserrat and Daza, Anamaría and Cabas-Mora, Gabriel and Sarmiento-Varón, Lindybeth and Sepúlveda-Yañez, Julieta and Anvari-Kazemabad, Hoda and Davari, Mehdi D and Uribe-Paredes, Roberto and Olivera-Nappa, \'Alvaro and Navarrete, Marcelo A and Medina-Ortiz, David},
    title = {Peptide-based drug discovery through artificial intelligence: towards an autonomous design of therapeutic peptides},
    journal = {Briefings in Bioinformatics},
    volume = {25},
    number = {4},
    pages = {bbae275},
    year = {2024},
    month = {06},
    issn = {1477-4054},
}

@article{fingerprints_vs_gnns,
author={Jiang, Dejun
and Wu, Zhenxing
and Hsieh, Chang-Yu
and Chen, Guangyong
and Liao, Ben
and Wang, Zhe
and Shen, Chao
and Cao, Dongsheng
and Wu, Jian
and Hou, Tingjun},
title={{Could graph neural networks learn better molecular representation for drug discovery? A comparison study of descriptor-based and graph-based models}},
journal={Journal of Cheminformatics},
year={2021},
month={Feb},
day={17},
volume={13},
number={1},
pages={12},
issn={1758-2946},
}

@article{ECFP,
  title={{Extended-Connectivity Fingerprints}},
  author={Rogers, David and Hahn, Mathew},
  journal={Journal of Chemical Information and Modeling},
  volume={50},
  number={5},
  pages={742--754},
  year={2010},
  publisher={ACS Publications}
}

@article{topological_torsion,
  title={{Topological torsion: a new molecular descriptor for SAR applications. Comparison with other descriptors}},
  author={Nilakantan, Ramaswamy and Bauman, Norman and Dixon, J Scott and Venkataraghavan, R},
  journal={Journal of Chemical Information and Computer Sciences},
  volume={27},
  number={2},
  pages={82--85},
  year={1987},
  publisher={ACS Publications}
}

@article{BERT_Protein,
    author = {Zhang, Yue and Lin, Jianyuan and Zhao, Lianmin and Zeng, Xiangxiang and Liu, Xiangrong},
    title = {{A novel antibacterial peptide recognition algorithm based on BERT}},
    journal = {Briefings in Bioinformatics},
    volume = {22},
    number = {6},
    pages = {bbab200},
    year = {2021},
    month = {05},
    issn = {1477-4054},
}

@inproceedings{GCN,
  title={{Semi-Supervised Classification with Graph Convolutional Networks}},
  author={Kipf, Thomas N. and Welling, Max},
  booktitle={International Conference on Learning Representations (ICLR)},
  year={2017}
}

@article{GraphSAGE,
  title={{Inductive Representation Learning on Large Graphs}},
  author={Hamilton, Will and Ying, Zhitao and Leskovec, Jure},
  journal={Advances in Neural Information Processing Systems},
  volume={30},
  year={2017}
}

@inproceedings{GIN,
title={{How Powerful are Graph Neural Networks?}},
author={Keyulu Xu and Weihua Hu and Jure Leskovec and Stefanie Jegelka},
booktitle={International Conference on Learning Representations},
year={2019},
}

@article{GNNs_as_continous_fingerprints,
  title={Convolutional networks on graphs for learning molecular fingerprints},
  author={Duvenaud, David K and Maclaurin, Dougal and Iparraguirre, Jorge and Bombarell, Rafael and Hirzel, Timothy and Aspuru-Guzik, Al{\'a}n and Adams, Ryan P},
  journal={Advances in Neural Information Processing Systems},
  volume={28},
  year={2015}
}

@book{molecular_descriptors,
  title={{Molecular descriptors for chemoinformatics: volume I: alphabetical listing/volume II: appendices, references}},
  author={Todeschini, Roberto and Consonni, Viviana},
  year={2009},
  publisher={John Wiley \& Sons}
}

@inproceedings{LightGBM,
 author = {Ke, Guolin and Meng, Qi and Finley, Thomas and Wang, Taifeng and Chen, Wei and Ma, Weidong and Ye, Qiwei and Liu, Tie-Yan},
 booktitle = {Advances in Neural Information Processing Systems},
 editor = {I. Guyon and U. Von Luxburg and S. Bengio and H. Wallach and R. Fergus and S. Vishwanathan and R. Garnett},
 pages = {},
 publisher = {Curran Associates, Inc.},
 title = {{LightGBM: A Highly Efficient Gradient Boosting Decision Tree}},
 volume = {30},
 year = {2017}
}

@misc{rdkit_fp,
  title = {{The RDKit Book: RDKit Fingerprints}},
  howpublished = {\url{https://www.rdkit.org/docs/RDKit_Book.html#rdkit-fingerprints}},
  note = {Accessed: 2025-01-24},
}

@article{scikit_fingerprints,
title = {{Scikit-fingerprints: Easy and efficient computation of molecular fingerprints in Python}},
journal = {SoftwareX},
volume = {28},
pages = {101944},
year = {2024},
issn = {2352-7110},
author = {Jakub Adamczyk and Piotr Ludynia},
}

@article{tunability_probst,
  author  = {Philipp Probst and Anne-Laure Boulesteix and Bernd Bischl},
  title   = {{Tunability: Importance of Hyperparameters of Machine Learning Algorithms}},
  journal = {Journal of Machine Learning Research},
  year    = {2019},
  volume  = {20},
  number  = {53},
  pages   = {1--32},
}

@article{virtual_screening,
author={Riniker, Sereina
and Landrum, Gregory A.},
title={{Open-source platform to benchmark fingerprints for ligand-based virtual screening}},
journal={Journal of Cheminformatics},
year={2013},
month={May},
day={30},
volume={5},
number={1},
pages={26},
issn={1758-2946},
}

@inproceedings{GraphViT,
  title={{A Generalization of ViT/MLP-Mixer to Graphs}},
  author={He, Xiaoxin and Hooi, Bryan and Laurent, Thomas and Perold, Adam and LeCun, Yann and Bresson, Xavier},
  booktitle={International Conference on Machine Learning},
  pages={12724--12745},
  year={2023},
  organization={PMLR}
}

@inproceedings{GRIT,
  title={{Graph Inductive Biases in Transformers without Message Passing}},
  author={Ma, Liheng and Lin, Chen and Lim, Derek and Romero-Soriano, Adriana and Dokania, Puneet K and Coates, Mark and Torr, Philip and Lim, Ser-Nam},
  booktitle={International Conference on Machine Learning},
  pages={23321--23337},
  year={2023},
  organization={PMLR}
}

@inproceedings{HDSE,
title={{Enhancing Graph Transformers with Hierarchical Distance Structural Encoding}},
author={Yuankai Luo and Hongkang Li and Lei Shi and Xiao-Ming Wu},
booktitle={The Thirty-eighth Annual Conference on Neural Information Processing Systems},
year={2024},
}

@article{LRGB,
  title={{Long Range Graph Benchmark}},
  author={Dwivedi, Vijay Prakash and Ramp{\'a}{\v{s}}ek, Ladislav and Galkin, Michael and Parviz, Ali and Wolf, Guy and Luu, Anh Tuan and Beaini, Dominique},
  journal={Advances in Neural Information Processing Systems},
  volume={35},
  pages={22326--22340},
  year={2022}
}

@article{MCC,
  title={The advantages of the Matthews correlation coefficient (MCC) over F1 score and accuracy in binary classification evaluation},
  author={Chicco, Davide and Jurman, Giuseppe},
  journal={BMC genomics},
  volume={21},
  number={1},
  pages={6},
  year={2020},
  publisher={Springer}
}

@incollection{MOLTOP,
  title={{Molecular Topological Profile (MOLTOP)-Simple and Strong Baseline for Molecular Graph Classification}},
  author={Adamczyk, Jakub and Czech, Wojciech},
  booktitle={ECAI 2024},
  pages={1575--1582},
  year={2024},
  publisher={IOS Press}
}

@article{LRGB_reassessment,
title={{Where Did the Gap Go? Reassessing the Long-Range Graph Benchmark}},
author={Jan T{\"o}nshoff and Martin Ritzert and Eran Rosenbluth and Martin Grohe},
journal={Transactions on Machine Learning Research},
issn={2835-8856},
year={2024},
}

@article{peptides_structure,
  title={Structural information in therapeutic peptides: Emerging applications in biomedicine},
  author={Iglesias, Valent{\'\i}n and B{\'a}rcenas, Oriol and Pintado-Grima, Carlos and Burdukiewicz, Micha{\l} and Ventura, Salvador},
  journal={FEBS Open Bio},
  volume={15},
  number={2},
  pages={254--268},
  year={2025},
  publisher={Wiley Online Library}
}

@article{rampavsek2022recipe,
  title={{Recipe for a General, Powerful, Scalable Graph Transformer}},
  author={Ramp{\'a}{\v{s}}ek, Ladislav and Galkin, Michael and Dwivedi, Vijay Prakash and Luu, Anh Tuan and Wolf, Guy and Beaini, Dominique},
  journal={Advances in Neural Information Processing Systems},
  volume={35},
  pages={14501--14515},
  year={2022}
}

@inproceedings{S2GCN,
title={{Spatio-Spectral Graph Neural Networks}},
author={Simon Geisler and Arthur Kosmala and Daniel Herbst and Stephan G{\"u}nnemann},
booktitle={The Thirty-eighth Annual Conference on Neural Information Processing Systems},
year={2024},
}

@article{lee2023amp,
  title={{AMP-BERT: Prediction of antimicrobial peptide function based on a BERT model}},
  author={Lee, Hansol and Lee, Songyeon and Lee, Ingoo and Nam, Hojung},
  journal={Protein Science},
  volume={32},
  number={1},
  pages={e4529},
  year={2023},
  publisher={Wiley Online Library}
}

@article{dee2022lmpred,
  title={{LMPred: Predicting antimicrobial peptides using pre-trained language models and deep learning}},
  author={Dee, William},
  journal={Bioinformatics Advances},
  volume={2},
  number={1},
  pages={vbac021},
  year={2022},
  publisher={Oxford University Press}
}

@article{ma2022identification,
  title={{Identification of antimicrobial peptides from the human gut microbiome using deep learning}},
  author={Ma, Yue and Guo, Zhengyan and Xia, Binbin and Zhang, Yuwei and Liu, Xiaolin and Yu, Ying and Tang, Na and Tong, Xiaomei and Wang, Min and Ye, Xin and others},
  journal={Nature Biotechnology},
  volume={40},
  number={6},
  pages={921--931},
  year={2022},
  publisher={Nature Publishing Group US New York}
}

@article{lawrence2021ampeppy,
  title={{amPEPpy 1.0: a portable and accurate antimicrobial peptide prediction tool}},
  author={Lawrence, Travis J and Carper, Dana L and Spangler, Margaret K and Carrell, Alyssa A and Rush, Tom{\'a}s A and Minter, Stephen J and Weston, David J and Labb{\'e}, Jessy L},
  journal={Bioinformatics},
  volume={37},
  number={14},
  pages={2058--2060},
  year={2021},
  publisher={Oxford University Press}
}

@article{veltri2018deep,
  title={{Deep learning improves antimicrobial peptide recognition}},
  author={Veltri, Daniel and Kamath, Uday and Shehu, Amarda},
  journal={Bioinformatics},
  volume={34},
  number={16},
  pages={2740--2747},
  year={2018},
  publisher={Oxford University Press}
}

@article{MACREL,
author = {Santos-Júnior, Célio Dias and Pan, Shaojun and Zhao, Xing-Ming and Coelho, Luis Pedro},
address = {United States},
copyright = {2020 Santos-Júnior et al.},
issn = {2167-8359},
journal = {PeerJ (San Francisco, CA)},
language = {eng},
pages = {e10555-e10555},
publisher = {PeerJ. Ltd},
title = {{Macrel: antimicrobial peptide screening in genomes and metagenomes}},
volume = {8},
year = {2020},
}

@article{AMPBenchmark,
    author = {Sidorczuk, Katarzyna and Gagat, Przemysław and Pietluch, Filip and Kała, Jakub and Rafacz, Dominik and Bąkała, Laura and Słowik, Jadwiga and Kolenda, Rafał and Rödiger, Stefan and Fingerhut, Legana C H W and Cooke, Ira R and Mackiewicz, Paweł and Burdukiewicz, Michał},
    title = {{Benchmarks in antimicrobial peptide prediction are biased due to the selection of negative data}},
    journal = {Briefings in Bioinformatics},
    volume = {23},
    number = {5},
    pages = {bbac343},
    year = {2022},
    month = {08},
    issn = {1477-4054},
}

@article{AMPfun,
    author = {Chung, Chia-Ru and Kuo, Ting-Rung and Wu, Li-Ching and Lee, Tzong-Yi and Horng, Jorng-Tzong},
    title = {{Characterization and identification of antimicrobial peptides with different functional activities}},
    journal = {Briefings in Bioinformatics},
    volume = {21},
    number = {3},
    pages = {1098-1114},
    year = {2019},
    month = {06},
    issn = {1477-4054},
}

@article{AmpGram,
AUTHOR = {Burdukiewicz, Michał and Sidorczuk, Katarzyna and Rafacz, Dominik and Pietluch, Filip and Chilimoniuk, Jarosław and Rödiger, Stefan and Gagat, Przemysław},
TITLE = {{Proteomic Screening for Prediction and Design of Antimicrobial Peptides with AmpGram}},
JOURNAL = {International Journal of Molecular Sciences},
VOLUME = {21},
YEAR = {2020},
NUMBER = {12},
ARTICLE-NUMBER = {4310},
PubMedID = {32560350},
ISSN = {1422-0067},
}

@article{ampir,
    author = {Fingerhut, Legana C H W and Miller, David J and Strugnell, Jan M and Daly, Norelle L and Cooke, Ira R},
    title = {{ampir: an R package for fast genome-wide prediction of antimicrobial peptides}},
    journal = {Bioinformatics},
    volume = {36},
    number = {21},
    pages = {5262-5263},
    year = {2020},
    month = {07},
    issn = {1367-4803},
}

@article{AutoPeptideML,
    author = {Fernández-Díaz, Raúl and Cossio-Pérez, Rodrigo and Agoni, Clement and Lam, Hoang Thanh and Lopez, Vanessa and Shields, Denis C},
    title = {{AutoPeptideML: a study on how to build more trustworthy peptide bioactivity predictors}},
    journal = {Bioinformatics},
    volume = {40},
    number = {9},
    pages = {btae555},
    year = {2024},
    month = {09},
    issn = {1367-4811},
}

@article{BERT_benchmark,
author = {Gao, Wanling and Zhao, Jun and Gui, Jianfeng and Wang, Zehan and Chen, Jie and Yue, Zhenyu},
title = {{Comprehensive Assessment of BERT-Based Methods for Predicting Antimicrobial Peptides}},
journal = {Journal of Chemical Information and Modeling},
volume = {64},
number = {19},
pages = {7772-7785},
year = {2024},
note ={PMID: 39316765},
}

@article{CDHIT,
    author = {Fu, Limin and Niu, Beifang and Zhu, Zhengwei and Wu, Sitao and Li, Weizhong},
    title = {{CD-HIT: accelerated for clustering the next-generation sequencing data}},
    journal = {Bioinformatics},
    volume = {28},
    number = {23},
    pages = {3150-3152},
    year = {2012},
    month = {10},
    issn = {1367-4803},
}

@Article{DeepAmPEP30,
author={Yan, Jielu
and Bhadra, Pratiti
and Li, Ang
and Sethiya, Pooja
and Qin, Longguang
and Tai, Hio Kuan
and Wong, Koon Ho
and Siu, Shirley W.I.},
title={{Deep-AmPEP30: Improve Short Antimicrobial Peptides Prediction with Deep Learning}},
journal={Molecular Therapy Nucleic Acids},
year={2020},
month={Jun},
day={05},
publisher={Elsevier},
volume={20},
pages={882-894},
issn={2162-2531},
}

@article{ESM2,
author = {Zeming Lin  and Halil Akin  and Roshan Rao  and Brian Hie  and Zhongkai Zhu  and Wenting Lu  and Nikita Smetanin  and Robert Verkuil  and Ori Kabeli  and Yaniv Shmueli  and Allan dos Santos Costa  and Maryam Fazel-Zarandi  and Tom Sercu  and Salvatore Candido  and Alexander Rives },
title = {{Evolutionary-scale prediction of atomic-level protein structure with a language model}},
journal = {Science},
volume = {379},
number = {6637},
pages = {1123-1130},
year = {2023},
}

@article{PeptideReactor,
    author = {Spänig, Sebastian and Mohsen, Siba and Hattab, Georges and Hauschild, Anne-Christin and Heider, Dominik},
    title = {{A large-scale comparative study on peptide encodings for biomedical classification}},
    journal = {NAR Genomics and Bioinformatics},
    volume = {3},
    number = {2},
    pages = {lqab039},
    year = {2021},
    month = {05},
    issn = {2631-9268},
}

@article{ProtTrans,
  title={{ProtTrans: Toward Understanding the Language of Life Through Self-Supervised Learning}},
  author={Elnaggar, Ahmed and Heinzinger, Michael and Dallago, Christian and Rehawi, Ghalia and Wang, Yu and Jones, Llion and Gibbs, Tom and Feher, Tamas and Angerer, Christoph and Steinegger, Martin and others},
  journal={IEEE Transactions on Pattern Analysis \& Machine Intelligence},
  volume={44},
  number={10},
  pages={7112--7127},
  year={2021},
  publisher={IEEE}
}

@article{Xu_AMP,
    author = {Xu, Jing and Li, Fuyi and Leier, André and Xiang, Dongxu and Shen, Hsin-Hui and Marquez Lago, Tatiana T and Li, Jian and Yu, Dong-Jun and Song, Jiangning},
    title = {{Comprehensive assessment of machine learning-based methods for predicting antimicrobial peptides}},
    journal = {Briefings in Bioinformatics},
    volume = {22},
    number = {5},
    pages = {bbab083},
    year = {2021},
    month = {03},
    issn = {1477-4054},
}

@article{extremely_randomized_trees,
  title={{Extremely randomized trees}},
  author={Geurts, Pierre and Ernst, Damien and Wehenkel, Louis},
  journal={Machine Learning},
  volume={63},
  pages={3--42},
  year={2006},
  publisher={Springer}
}

@article{sequence_motif_KKK,
  title={Design and synthesis of a potent peptide containing both specific and non-specific cell-adhesion motifs},
  author={Lai, Yuxiao and Xie, Cao and Zhang, Zheng and Lu, Weiyue and Ding, Jiandong},
  journal={Biomaterials},
  volume={31},
  number={18},
  pages={4809--4817},
  year={2010},
  publisher={Elsevier}
}

@article{sequence_motif_KKK_2,
  title={A 5-lipoxygenase-specific sequence motif impedes enzyme activity and confers dependence on a partner protein},
  author={Schexnaydre, Erin E and Gerstmeier, Jana and Garscha, Ulrike and Jordan, Paul M and Werz, Oliver and Newcomer, Marcia E},
  journal={Biochimica et Biophysica Acta (BBA)-Molecular and Cell Biology of Lipids},
  volume={1864},
  number={4},
  pages={543--551},
  year={2019},
  publisher={Elsevier}
}

@article{sequence_motif_RRR,
  title={Short, mirror-symmetric antimicrobial peptides centered on “RRR” have broad-spectrum antibacterial activity with low drug resistance and toxicity},
  author={Zhang, Fangyan and Yang, Ping and Mao, Wenbo and Zhong, Chao and Zhang, Jingying and Chang, Linlin and Wu, Xiaoyan and Liu, Hui and Zhang, Yun and Gou, Sanhu and others},
  journal={Acta Biomaterialia},
  volume={154},
  pages={145--167},
  year={2022},
  publisher={Elsevier}
}

\clearpage

\onecolumn

\begin{center}
    \textbf{\LARGE Supplementary information}
\end{center}

\section{Reproducibility and hardware details}
\label{appendix_reproducibility}

To ensure the full reproducibility of our results, we used \texttt{uv} to pin the exact version of all dependencies in the project, including transitive dependencies of directly used libraries. We distribute the \texttt{pyproject.toml} and resulting \texttt{uv.lock} file with our source code. This ensures the exact reproducibility of all results that is OS-agnostic and hardware-agnostic.

We conduct experiments primarily on CPU, as our method does not require GPU. Only for ESM2 we used NVidia RTX 3080 GPU with 12 GB VRAM. All experiments were run on a machine with Intel Core i7-12700KF 3.61 GHz CPU and 96 GB RAM, running Linux Ubuntu 24.04 OS. We additionally ran the experiments on a second machine with Intel Core i7-10850H 2.70 GHz CPU and 32 GB RAM, running Linux Ubuntu 22.04 OS. The results were exactly the same in all cases.

\section{Targets in specific benchmarks}
\label{benchmark_descriptions}

Here we describe in detail targets of each benchmark.

\subsection{AMPBenchmark}

In AMPBenchmark \cite{AMPBenchmark} authors focus on one task, which is antimicrobial activity. The dataset includes a test set consisting of 5 repeats and multiple training sets, each also consisting of 5 repeats. All training sets and their repeats use the same 4151 samples as their entire positive class, and all repeats of the test set use the same 1039 positive samples similarly. Repeats and training sets differ by the negatives sampling method, as described in the main body.

\begin{figure}[t]
    \centering
    \includegraphics[width=0.8\textwidth]{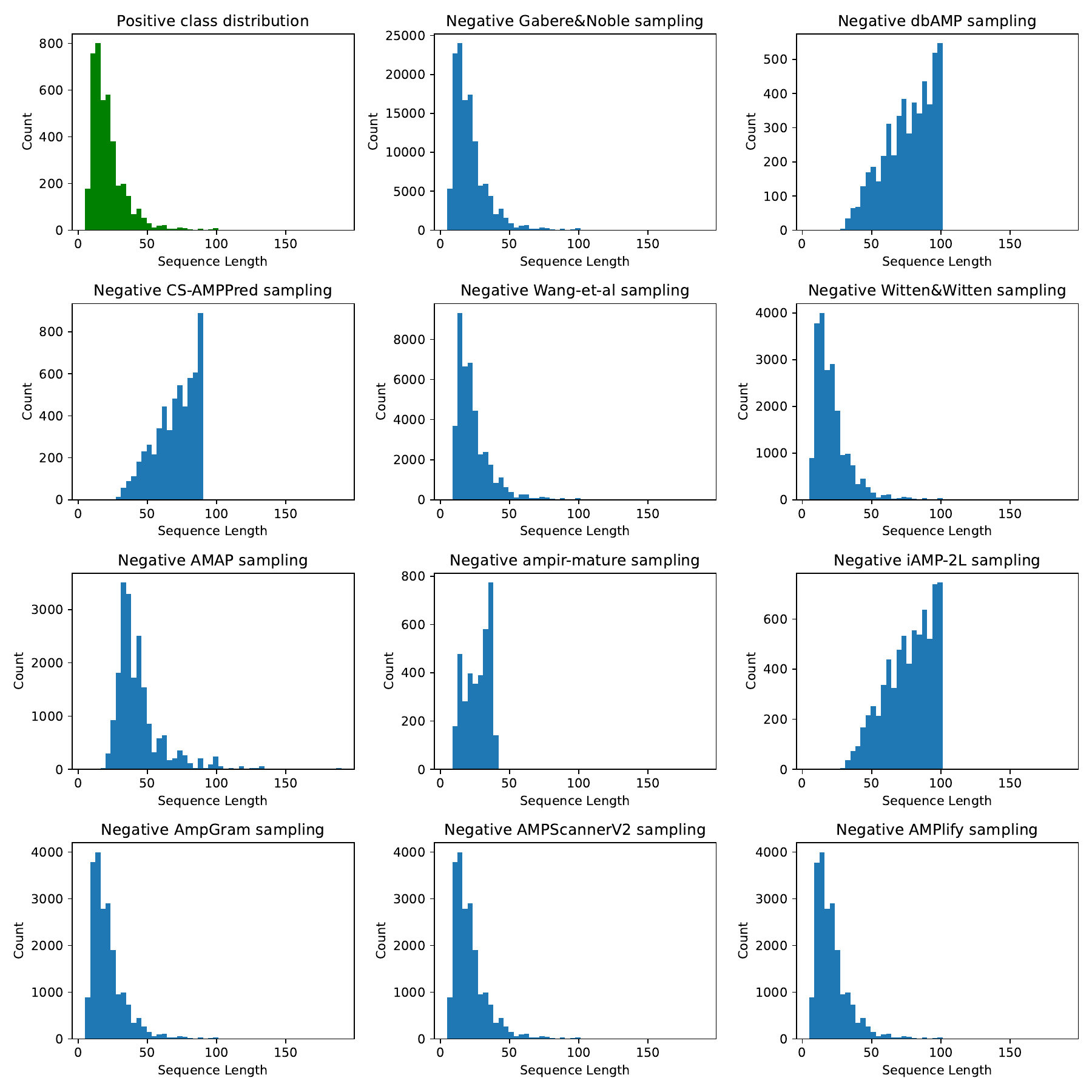}
    \caption{AMPBenchmark \cite{AMPBenchmark} sequence length distributions for different negative sampling methods.}
    \label{fig:ampbenchmark_sequence_length_distribution}
\end{figure}

Each training set is defined by a sampling method used to chose negative class samples. The samples vary by number and length distribution. We report the distributions of sequence length for positive class and for negative class generated by each method in Figure \ref{fig:ampbenchmark_sequence_length_distribution}. 
We also report the number of unique samples generated by each method and average number of common samples within two different repeats of the same method in Table \ref{appendix_ampbenchmark_class_counts}.

\begin{table}[th!]
\centering
\caption{AMPBenchmark negative class counts and average common sample count off different subsampling methods.}
\begin{tabular}{|l|l|l|}
\hline
\textbf{Sampling method} & \textbf{Unique samples} & \textbf{Avg. common samples count} \\ \hline
Gabere\&Noble   & 124504                & 1.7                    \\ \hline
dbAMP           & 5138                  & 3285.8                 \\ \hline
CS-AMPPred      & 5829                  & 2942.3                 \\ \hline
Wang-et-al      & 41784                 & 0.1                    \\ \hline
Witten\&Witten  & 20754                 & 0.1                    \\ \hline
AMAP            & 20046                 & 0                      \\ \hline
ampir-mature    & 3579                  & 2285                   \\ \hline
iAMP-2L         & 7327                  & 4678.3                 \\ \hline
AmpGram         & 20754                 & 0.1                    \\ \hline
AMPScannerV2    & 20755                 & 0                      \\ \hline
AMPlify         & 20752                 & 0.1                    \\ \hline
\end{tabular}
\label{appendix_ampbenchmark_class_counts}
\end{table}

\clearpage

\subsection{AutoPeptideML}

In AutoPeptideML \cite{AutoPeptideML}, authors use 16 independent tasks predicting targets to assess predictive performance against varied biological endpoints. Each classification task is almost perfectly balanced. These tasks include: antibacterial activity, ACE inhibition, anticancer activity, antifungal activity, antimalarial activity, antimicrobial activity, antioxidant activity, antiparasitic activity, antiviral activity, blood–brain barrier penetration, DPPIV inhibition, anti-MRSA activity, neuropeptide activity, quorum sensing activity, toxicity prediction, and tumor T-cell antigen identification.

\begin{figure}[th!]
    \centering
    \includegraphics[width=0.8\textwidth]{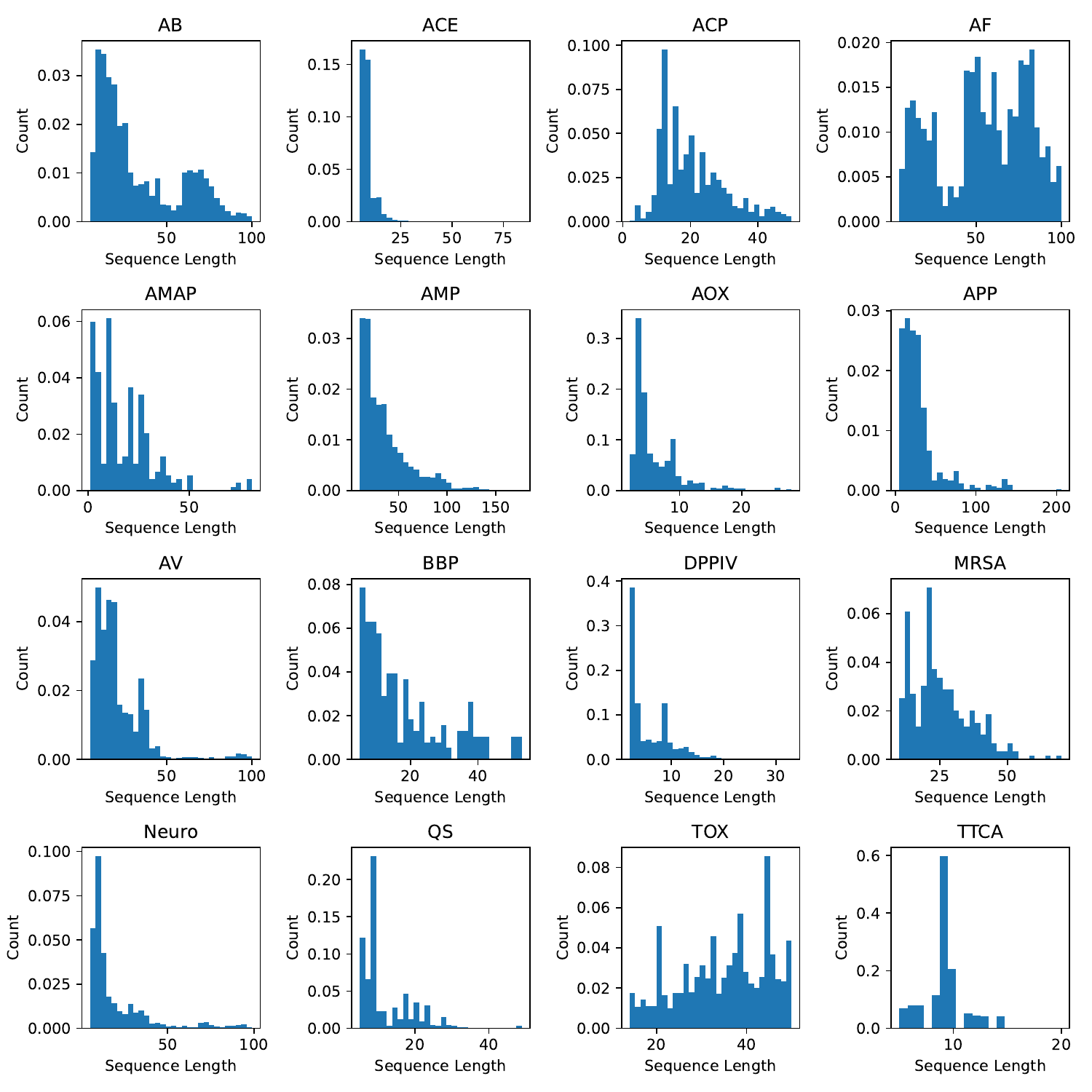}
    \caption{AutoPeptideML \cite{AutoPeptideML} sequence length distribution across different datasets.}
    \label{fig:autopeptideml_sequence_length_distribution}
\end{figure}

We report sequence length distributions for each task in Figure \ref{fig:autopeptideml_sequence_length_distribution}. We additionally report the number of train and test samples for each of the targets in Table \ref{appendix_autopeptideml_dataset_sizes}.

\begin{table}[th!]
\centering
\caption{Dataset sizes in AutoPeptideML \cite{AutoPeptideML}.}
\begin{tabular}{|l|l|l|}
\hline
\textbf{Target} & \textbf{Train size} & \textbf{Test size} \\ \hline
AB              & 13245               & 3311               \\ \hline
ACE             & 1685                & 421                \\ \hline
ACP             & 1378                & 344                \\ \hline
AF              & 1591                & 395                \\ \hline
AMAP            & 221                 & 55                 \\ \hline
AMP             & 10266               & 2566               \\ \hline
AOX             & 698                 & 174                \\ \hline
APP             & 482                 & 120                \\ \hline
AV              & 4711                & 1177               \\ \hline
BBP             & 191                 & 47                 \\ \hline
DPPIV           & 1063                & 265                \\ \hline
MRSA            & 237                 & 59                 \\ \hline
Neuro           & 3881                & 969                \\ \hline
QS              & 349                 & 87                 \\ \hline
TOX             & 3092                & 772                \\ \hline
TTCA            & 948                 & 236                \\ \hline
\end{tabular}

\label{appendix_autopeptideml_dataset_sizes}
\end{table}

\clearpage

\subsection{BERT-based models benchmark}

In BERT-based models benchmark \cite{BERT_benchmark} authors focus on antimicrobial peptides (AMPs) classification using 6 different datasets. As described in the main body, BERT models are pretrained, and whole datasets are used for testing in the aforementioned publication. To mimic this with our approach, we use a leave-one-dataset-out strategy. For example, when evaluating on the ADAPTABLE dataset, we merge all remaining datasets into a single training set and apply CD-HIT-2D \cite{CDHIT} at a $40\%$ threshold to remove peptides overly similar to those in the test set.

\begin{figure}[th!]
    \centering
    \includegraphics[width=0.8\textwidth]{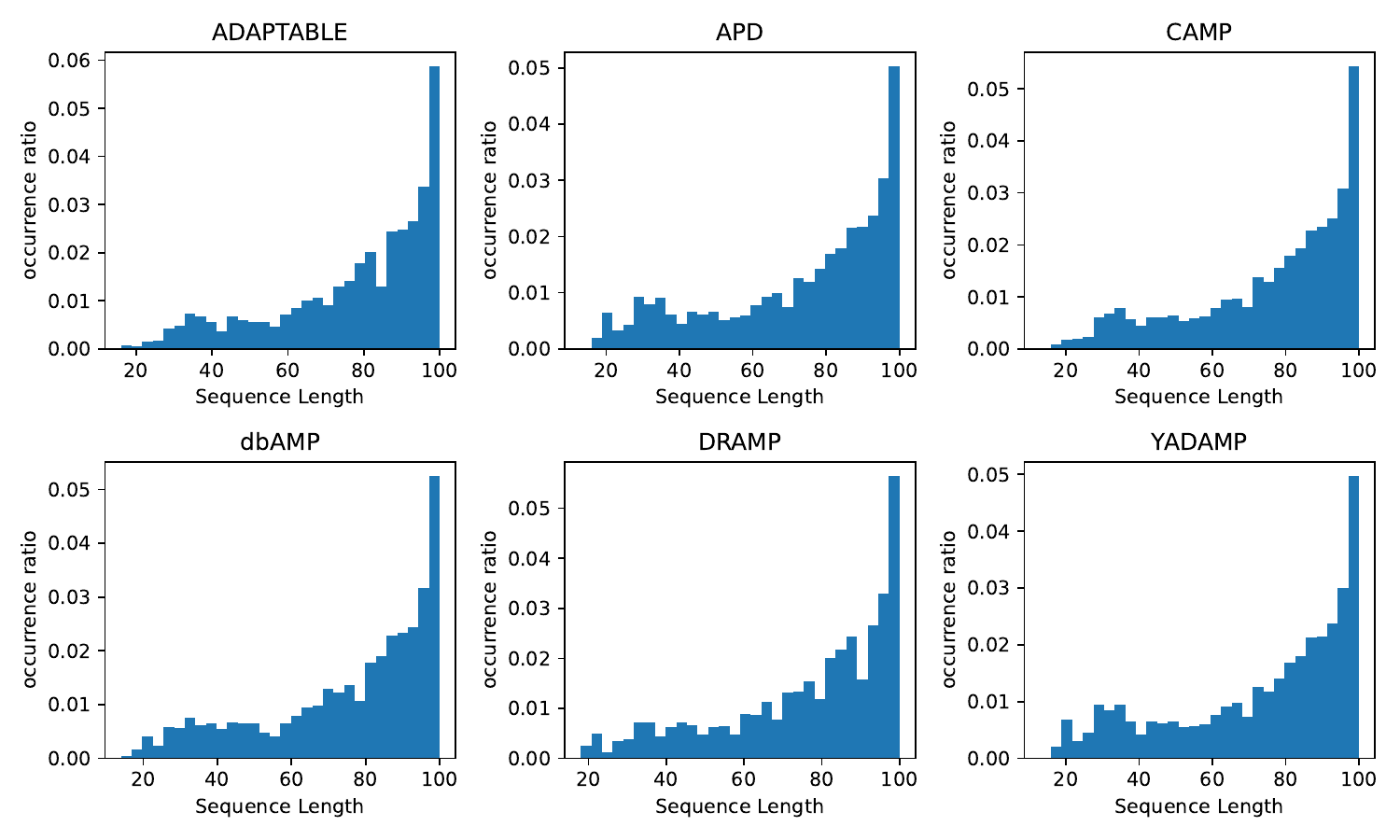}
    \caption{BERT AMP Benchmark \cite{BERT_benchmark} sequence length distribution across different datasets.}
    \label{fig:bert_amp_benchmark_sequence_length_distribution}
\end{figure}

This approach creates 6 new datasets, with the original one being the test set, and all others, after CD-HIT-2D, become the training set. We summarize the sequence length distributions of resulting datasets in Figure \ref{fig:bert_amp_benchmark_sequence_length_distribution}. We include the train and test set sizes, as well as percentage of positive class for each task, in Table \ref{appendix_bert_amp_benchmark_dataset_sizes}. Training sets have different positive class percentage due to the applied procedure, but test sets are exactly the same, original datasets, as used in \cite{BERT_benchmark}, and have 50\% negative and positive class.

\begin{table}[th!]
\centering
\caption{Dataset sizes in BERT-based models benchmark \cite{BERT_benchmark}.}
\resizebox{\textwidth}{!}{
\begin{tabular}{|l|l|l|l|l|}
\hline
\textbf{Dataset} & \textbf{Train size} & \textbf{Train positive class \%} & \textbf{Test size} & \textbf{Test positive class \%} \\ \hline
ADAPTABLE        & 2282                    & 36.33\%                          & 2170                   & 50.00\%                         \\ \hline
APD              & 5887                    & 49.26\%                          & 68                     & 50.00\%                         \\ \hline
CAMP             & 3773                    & 43.97\%                          & 1154                   & 50.00\%                         \\ \hline
dbAMP            & 4905                    & 47.46\%                          & 568                    & 50.00\%                         \\ \hline
DRAMP            & 2731                    & 45.26\%                          & 2082                   & 50.00\%                         \\ \hline
YADAMP           & 5978                    & 49.75\%                          & 46                     & 50.00\%                         \\ \hline
\end{tabular}
}
\label{appendix_bert_amp_benchmark_dataset_sizes}
\end{table}

\clearpage

\subsection{PeptideReactor}

PeptideReactor \cite{PeptideReactor} is a large benchmark including 50 binary classification datasets. Their tasks include A-cell epitope classification, three anticancer tasks, two antifungal tasks, two anti-inflammatory tasks, seven antimicrobial tasks, two antibacterial tasks, two antiviral tasks, linear B-cell epitope classification, eight cell-penetrating tasks, hemolytic activity classification, seventeen HIV-related tasks, immunosuppressive activity based on IL-10 induction, insect neuropeptide classification, proinflammatory activity classification, and T-cell epitope classification, all formulated as binary classification tasks.

\begin{figure}[th!]
    \centering
    \includegraphics[width=0.8\textwidth]{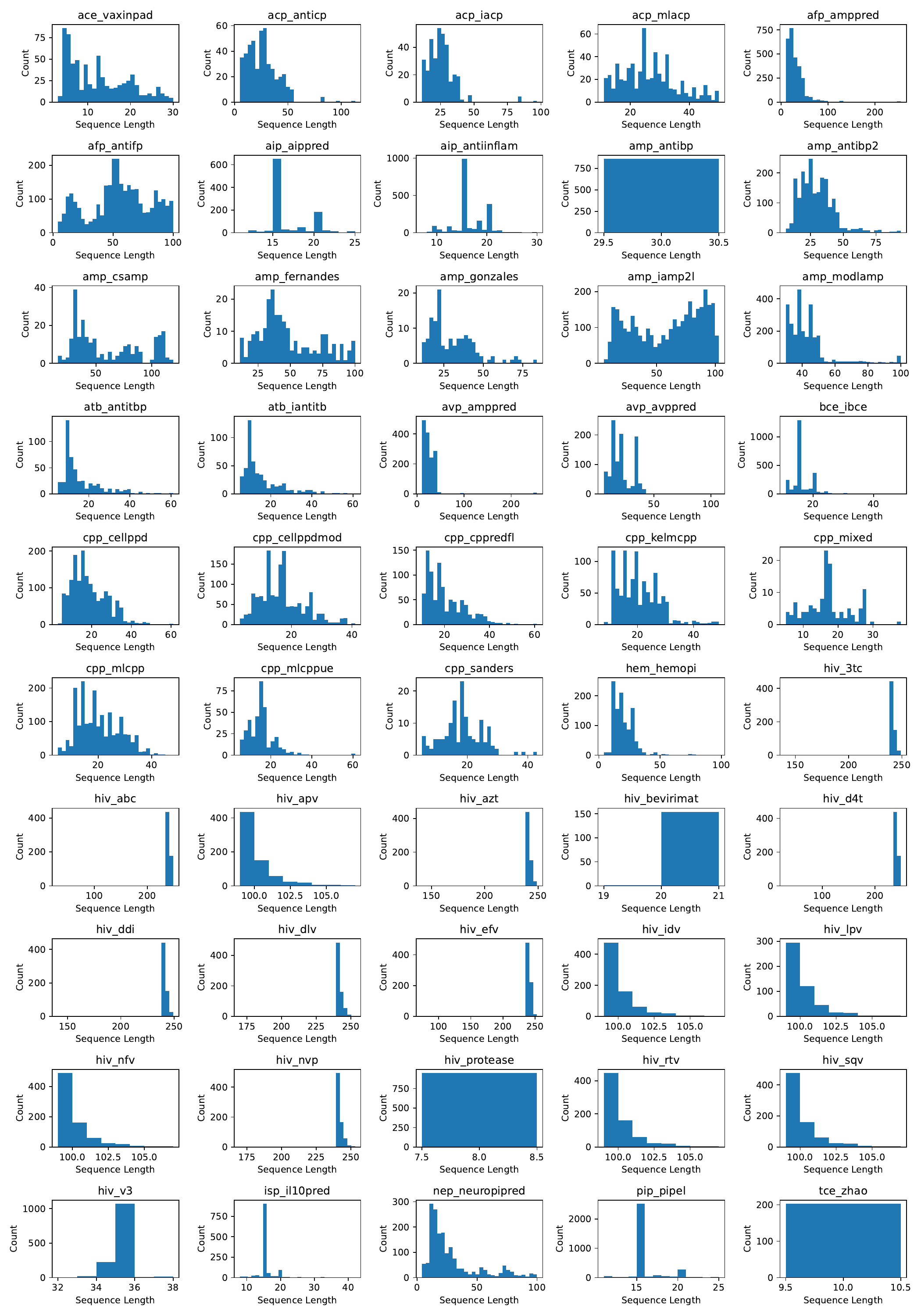}
    \caption{PeptideReactor \cite{PeptideReactor} sequence length distribution across different tasks.}
    \label{fig:peptidereactor_sequence_length_distribution}
\end{figure}

We report sequence length distribution for each of these tasks in Figure \ref{fig:peptidereactor_sequence_length_distribution}. It is worth noting that many of the tasks are fixed on a specific sequence length for the entire dataset. Train-test splits are not provided for the benchmark, as it uses 10 repetitions of 5-fold cross-validation. In Table \ref{appendix_peptidereactor_dataset_sizes}, we report: size of each dataset, positive class percentage, and statistics of sequence length distribution (not plots due to large number of datasets).

\begin{table}[]
\centering
\caption{Dataset sizes in PeptideReactor\cite{PeptideReactor}.}
\resizebox{0.8\textwidth}{!}{
\begin{tabular}{|c|c|c|ccccccc|}
\hline
\multirow{2}{*}{\textbf{Dataset}} & \multirow{2}{*}{\textbf{\begin{tabular}[c]{@{}c@{}}Dataset\\ size\end{tabular}}} & \multirow{2}{*}{\textbf{\begin{tabular}[c]{@{}c@{}}Positive\\ class \%\end{tabular}}} & \multicolumn{7}{c|}{\textbf{Sequence length}} \\ \cline{4-10} 
 &  &  & \multicolumn{1}{c|}{\textbf{min}} & \multicolumn{1}{c|}{\textbf{p5}} & \multicolumn{1}{c|}{\textbf{p25}} & \multicolumn{1}{c|}{\textbf{median}} & \multicolumn{1}{c|}{\textbf{p75}} & \multicolumn{1}{c|}{\textbf{p95}} & \textbf{max} \\ \hline
ace\_vaxinpad & 688 & 44.04\% & \multicolumn{1}{c|}{3} & \multicolumn{1}{c|}{4} & \multicolumn{1}{c|}{5} & \multicolumn{1}{c|}{10} & \multicolumn{1}{c|}{17} & \multicolumn{1}{c|}{26} & 30 \\ \hline
acp\_anticp & 450 & 50.00\% & \multicolumn{1}{c|}{5} & \multicolumn{1}{c|}{7} & \multicolumn{1}{c|}{15} & \multicolumn{1}{c|}{25} & \multicolumn{1}{c|}{34} & \multicolumn{1}{c|}{49} & 113 \\ \hline
acp\_iacp & 344 & 40.12\% & \multicolumn{1}{c|}{11} & \multicolumn{1}{c|}{13} & \multicolumn{1}{c|}{18} & \multicolumn{1}{c|}{25} & \multicolumn{1}{c|}{30} & \multicolumn{1}{c|}{38} & 97 \\ \hline
acp\_mlacp & 585 & 31.97\% & \multicolumn{1}{c|}{11} & \multicolumn{1}{c|}{13} & \multicolumn{1}{c|}{20} & \multicolumn{1}{c|}{25} & \multicolumn{1}{c|}{32} & \multicolumn{1}{c|}{45} & 50 \\ \hline
afp\_amppred & 2768 & 50.00\% & \multicolumn{1}{c|}{10} & \multicolumn{1}{c|}{13} & \multicolumn{1}{c|}{19} & \multicolumn{1}{c|}{26} & \multicolumn{1}{c|}{37} & \multicolumn{1}{c|}{67} & 255 \\ \hline
afp\_antifp & 2916 & 50.03\% & \multicolumn{1}{c|}{4} & \multicolumn{1}{c|}{12} & \multicolumn{1}{c|}{41} & \multicolumn{1}{c|}{55} & \multicolumn{1}{c|}{73} & \multicolumn{1}{c|}{94} & 100 \\ \hline
aip\_aippred & 1049 & 40.04\% & \multicolumn{1}{c|}{11} & \multicolumn{1}{c|}{15} & \multicolumn{1}{c|}{15} & \multicolumn{1}{c|}{15} & \multicolumn{1}{c|}{18} & \multicolumn{1}{c|}{20} & 25 \\ \hline
aip\_antiinflam & 2124 & 40.63\% & \multicolumn{1}{c|}{7} & \multicolumn{1}{c|}{9} & \multicolumn{1}{c|}{15} & \multicolumn{1}{c|}{15} & \multicolumn{1}{c|}{19} & \multicolumn{1}{c|}{21} & 30 \\ \hline
amp\_antibp & 861 & 50.06\% & \multicolumn{1}{c|}{30} & \multicolumn{1}{c|}{30} & \multicolumn{1}{c|}{30} & \multicolumn{1}{c|}{30} & \multicolumn{1}{c|}{30} & \multicolumn{1}{c|}{30} & 30 \\ \hline
amp\_antibp2 & 1993 & 50.13\% & \multicolumn{1}{c|}{6} & \multicolumn{1}{c|}{13} & \multicolumn{1}{c|}{20} & \multicolumn{1}{c|}{27} & \multicolumn{1}{c|}{37} & \multicolumn{1}{c|}{53} & 94 \\ \hline
amp\_csamp & 256 & 50.00\% & \multicolumn{1}{c|}{16} & \multicolumn{1}{c|}{29} & \multicolumn{1}{c|}{34} & \multicolumn{1}{c|}{47} & \multicolumn{1}{c|}{85} & \multicolumn{1}{c|}{110} & 119 \\ \hline
amp\_fernandes & 231 & 49.78\% & \multicolumn{1}{c|}{11} & \multicolumn{1}{c|}{18} & \multicolumn{1}{c|}{32} & \multicolumn{1}{c|}{41} & \multicolumn{1}{c|}{63} & \multicolumn{1}{c|}{92} & 100 \\ \hline
amp\_gonzales & 129 & 20.93\% & \multicolumn{1}{c|}{11} & \multicolumn{1}{c|}{14} & \multicolumn{1}{c|}{20} & \multicolumn{1}{c|}{26} & \multicolumn{1}{c|}{38} & \multicolumn{1}{c|}{55} & 84 \\ \hline
amp\_iamp2l & 3284 & 26.77\% & \multicolumn{1}{c|}{5} & \multicolumn{1}{c|}{13} & \multicolumn{1}{c|}{32} & \multicolumn{1}{c|}{65} & \multicolumn{1}{c|}{86} & \multicolumn{1}{c|}{98} & 103 \\ \hline
amp\_modlamp & 2579 & 47.50\% & \multicolumn{1}{c|}{30} & \multicolumn{1}{c|}{30} & \multicolumn{1}{c|}{35} & \multicolumn{1}{c|}{40} & \multicolumn{1}{c|}{46} & \multicolumn{1}{c|}{73} & 100 \\ \hline
atb\_antitbp & 492 & 50.00\% & \multicolumn{1}{c|}{5} & \multicolumn{1}{c|}{7} & \multicolumn{1}{c|}{9} & \multicolumn{1}{c|}{12} & \multicolumn{1}{c|}{20} & \multicolumn{1}{c|}{38} & 61 \\ \hline
atb\_iantitb & 492 & 50.00\% & \multicolumn{1}{c|}{5} & \multicolumn{1}{c|}{6} & \multicolumn{1}{c|}{9} & \multicolumn{1}{c|}{12} & \multicolumn{1}{c|}{19} & \multicolumn{1}{c|}{38} & 61 \\ \hline
avp\_amppred & 1478 & 50.00\% & \multicolumn{1}{c|}{10} & \multicolumn{1}{c|}{12} & \multicolumn{1}{c|}{17} & \multicolumn{1}{c|}{22} & \multicolumn{1}{c|}{34} & \multicolumn{1}{c|}{39} & 255 \\ \hline
avp\_avppred & 1047 & 57.21\% & \multicolumn{1}{c|}{6} & \multicolumn{1}{c|}{8} & \multicolumn{1}{c|}{15} & \multicolumn{1}{c|}{20} & \multicolumn{1}{c|}{30} & \multicolumn{1}{c|}{37} & 107 \\ \hline
bce\_ibce & 2518 & 44.08\% & \multicolumn{1}{c|}{11} & \multicolumn{1}{c|}{12} & \multicolumn{1}{c|}{15} & \multicolumn{1}{c|}{15} & \multicolumn{1}{c|}{18} & \multicolumn{1}{c|}{22} & 49 \\ \hline
cpp\_cellppd & 1614 & 50.00\% & \multicolumn{1}{c|}{3} & \multicolumn{1}{c|}{6} & \multicolumn{1}{c|}{12} & \multicolumn{1}{c|}{17} & \multicolumn{1}{c|}{25} & \multicolumn{1}{c|}{34} & 61 \\ \hline
cpp\_cellppdmod & 1462 & 50.07\% & \multicolumn{1}{c|}{3} & \multicolumn{1}{c|}{7} & \multicolumn{1}{c|}{12} & \multicolumn{1}{c|}{16} & \multicolumn{1}{c|}{21} & \multicolumn{1}{c|}{30} & 41 \\ \hline
cpp\_cppredfl & 924 & 50.00\% & \multicolumn{1}{c|}{10} & \multicolumn{1}{c|}{11} & \multicolumn{1}{c|}{14} & \multicolumn{1}{c|}{18} & \multicolumn{1}{c|}{26} & \multicolumn{1}{c|}{37} & 61 \\ \hline
cpp\_kelmcpp & 1003 & 50.25\% & \multicolumn{1}{c|}{8} & \multicolumn{1}{c|}{12} & \multicolumn{1}{c|}{15} & \multicolumn{1}{c|}{19} & \multicolumn{1}{c|}{25} & \multicolumn{1}{c|}{30} & 49 \\ \hline
cpp\_mixed & 128 & 75.78\% & \multicolumn{1}{c|}{5} & \multicolumn{1}{c|}{7} & \multicolumn{1}{c|}{13} & \multicolumn{1}{c|}{16} & \multicolumn{1}{c|}{21} & \multicolumn{1}{c|}{27} & 38 \\ \hline
cpp\_mlcpp & 1903 & 38.78\% & \multicolumn{1}{c|}{5} & \multicolumn{1}{c|}{10} & \multicolumn{1}{c|}{14} & \multicolumn{1}{c|}{19} & \multicolumn{1}{c|}{26} & \multicolumn{1}{c|}{34} & 48 \\ \hline
cpp\_mlcppue & 374 & 50.00\% & \multicolumn{1}{c|}{5} & \multicolumn{1}{c|}{7} & \multicolumn{1}{c|}{11} & \multicolumn{1}{c|}{15} & \multicolumn{1}{c|}{18} & \multicolumn{1}{c|}{27} & 61 \\ \hline
cpp\_sanders & 145 & 76.55\% & \multicolumn{1}{c|}{5} & \multicolumn{1}{c|}{7} & \multicolumn{1}{c|}{15} & \multicolumn{1}{c|}{18} & \multicolumn{1}{c|}{22} & \multicolumn{1}{c|}{27} & 43 \\ \hline
hem\_hemopi & 1104 & 47.28\% & \multicolumn{1}{c|}{4} & \multicolumn{1}{c|}{11} & \multicolumn{1}{c|}{14} & \multicolumn{1}{c|}{18} & \multicolumn{1}{c|}{25} & \multicolumn{1}{c|}{34} & 98 \\ \hline
hiv\_3tc & 624 & 31.25\% & \multicolumn{1}{c|}{141} & \multicolumn{1}{c|}{240} & \multicolumn{1}{c|}{240} & \multicolumn{1}{c|}{241} & \multicolumn{1}{c|}{242} & \multicolumn{1}{c|}{245} & 249 \\ \hline
hiv\_abc & 619 & 28.92\% & \multicolumn{1}{c|}{31} & \multicolumn{1}{c|}{240} & \multicolumn{1}{c|}{240} & \multicolumn{1}{c|}{241} & \multicolumn{1}{c|}{242} & \multicolumn{1}{c|}{245} & 249 \\ \hline
hiv\_apv & 702 & 60.40\% & \multicolumn{1}{c|}{99} & \multicolumn{1}{c|}{99} & \multicolumn{1}{c|}{99} & \multicolumn{1}{c|}{99} & \multicolumn{1}{c|}{100} & \multicolumn{1}{c|}{102} & 107 \\ \hline
hiv\_azt & 621 & 51.85\% & \multicolumn{1}{c|}{141} & \multicolumn{1}{c|}{240} & \multicolumn{1}{c|}{240} & \multicolumn{1}{c|}{241} & \multicolumn{1}{c|}{242} & \multicolumn{1}{c|}{245} & 249 \\ \hline
hiv\_bevirimat & 155 & 27.74\% & \multicolumn{1}{c|}{19} & \multicolumn{1}{c|}{20} & \multicolumn{1}{c|}{21} & \multicolumn{1}{c|}{21} & \multicolumn{1}{c|}{21} & \multicolumn{1}{c|}{21} & 21 \\ \hline
hiv\_d4t & 621 & 54.11\% & \multicolumn{1}{c|}{31} & \multicolumn{1}{c|}{240} & \multicolumn{1}{c|}{240} & \multicolumn{1}{c|}{241} & \multicolumn{1}{c|}{242} & \multicolumn{1}{c|}{245} & 249 \\ \hline
hiv\_ddi & 623 & 49.12\% & \multicolumn{1}{c|}{141} & \multicolumn{1}{c|}{240} & \multicolumn{1}{c|}{240} & \multicolumn{1}{c|}{241} & \multicolumn{1}{c|}{242} & \multicolumn{1}{c|}{245} & 249 \\ \hline
hiv\_dlv & 718 & 63.37\% & \multicolumn{1}{c|}{170} & \multicolumn{1}{c|}{240} & \multicolumn{1}{c|}{240} & \multicolumn{1}{c|}{241} & \multicolumn{1}{c|}{242} & \multicolumn{1}{c|}{246} & 253 \\ \hline
hiv\_efv & 721 & 62.00\% & \multicolumn{1}{c|}{74} & \multicolumn{1}{c|}{240} & \multicolumn{1}{c|}{240} & \multicolumn{1}{c|}{241} & \multicolumn{1}{c|}{242} & \multicolumn{1}{c|}{246} & 253 \\ \hline
hiv\_idv & 758 & 50.66\% & \multicolumn{1}{c|}{99} & \multicolumn{1}{c|}{99} & \multicolumn{1}{c|}{99} & \multicolumn{1}{c|}{99} & \multicolumn{1}{c|}{100} & \multicolumn{1}{c|}{102} & 107 \\ \hline
hiv\_lpv & 501 & 44.51\% & \multicolumn{1}{c|}{99} & \multicolumn{1}{c|}{99} & \multicolumn{1}{c|}{99} & \multicolumn{1}{c|}{99} & \multicolumn{1}{c|}{100} & \multicolumn{1}{c|}{102} & 107 \\ \hline
hiv\_nfv & 775 & 39.10\% & \multicolumn{1}{c|}{99} & \multicolumn{1}{c|}{99} & \multicolumn{1}{c|}{99} & \multicolumn{1}{c|}{99} & \multicolumn{1}{c|}{100} & \multicolumn{1}{c|}{102} & 107 \\ \hline
hiv\_nvp & 733 & 56.62\% & \multicolumn{1}{c|}{170} & \multicolumn{1}{c|}{240} & \multicolumn{1}{c|}{240} & \multicolumn{1}{c|}{241} & \multicolumn{1}{c|}{242} & \multicolumn{1}{c|}{246} & 253 \\ \hline
hiv\_protease & 947 & 15.73\% & \multicolumn{1}{c|}{8} & \multicolumn{1}{c|}{8} & \multicolumn{1}{c|}{8} & \multicolumn{1}{c|}{8} & \multicolumn{1}{c|}{8} & \multicolumn{1}{c|}{8} & 8 \\ \hline
hiv\_rtv & 728 & 47.94\% & \multicolumn{1}{c|}{99} & \multicolumn{1}{c|}{99} & \multicolumn{1}{c|}{99} & \multicolumn{1}{c|}{99} & \multicolumn{1}{c|}{100} & \multicolumn{1}{c|}{102} & 107 \\ \hline
hiv\_sqv & 761 & 60.05\% & \multicolumn{1}{c|}{99} & \multicolumn{1}{c|}{99} & \multicolumn{1}{c|}{99} & \multicolumn{1}{c|}{99} & \multicolumn{1}{c|}{100} & \multicolumn{1}{c|}{102} & 107 \\ \hline
hiv\_v3 & 1351 & 14.80\% & \multicolumn{1}{c|}{32} & \multicolumn{1}{c|}{34} & \multicolumn{1}{c|}{35} & \multicolumn{1}{c|}{35} & \multicolumn{1}{c|}{35} & \multicolumn{1}{c|}{35} & 38 \\ \hline
isp\_il10pred & 1242 & 31.72\% & \multicolumn{1}{c|}{8} & \multicolumn{1}{c|}{13} & \multicolumn{1}{c|}{15} & \multicolumn{1}{c|}{15} & \multicolumn{1}{c|}{15} & \multicolumn{1}{c|}{20} & 42 \\ \hline
nep\_neuropipred & 1750 & 50.00\% & \multicolumn{1}{c|}{4} & \multicolumn{1}{c|}{10} & \multicolumn{1}{c|}{14} & \multicolumn{1}{c|}{20} & \multicolumn{1}{c|}{31} & \multicolumn{1}{c|}{77} & 100 \\ \hline
pip\_pipel & 3228 & 25.81\% & \multicolumn{1}{c|}{11} & \multicolumn{1}{c|}{15} & \multicolumn{1}{c|}{15} & \multicolumn{1}{c|}{15} & \multicolumn{1}{c|}{15} & \multicolumn{1}{c|}{20} & 25 \\ \hline
tce\_zhao & 203 & 17.73\% & \multicolumn{1}{c|}{10} & \multicolumn{1}{c|}{10} & \multicolumn{1}{c|}{10} & \multicolumn{1}{c|}{10} & \multicolumn{1}{c|}{10} & \multicolumn{1}{c|}{10} & 10 \\ \hline
\end{tabular}
}
\label{appendix_peptidereactor_dataset_sizes}
\end{table}

\clearpage

\subsection{XUAMP}

\begin{figure}[th!]
    \centering
    \includegraphics[width=0.8\textwidth]{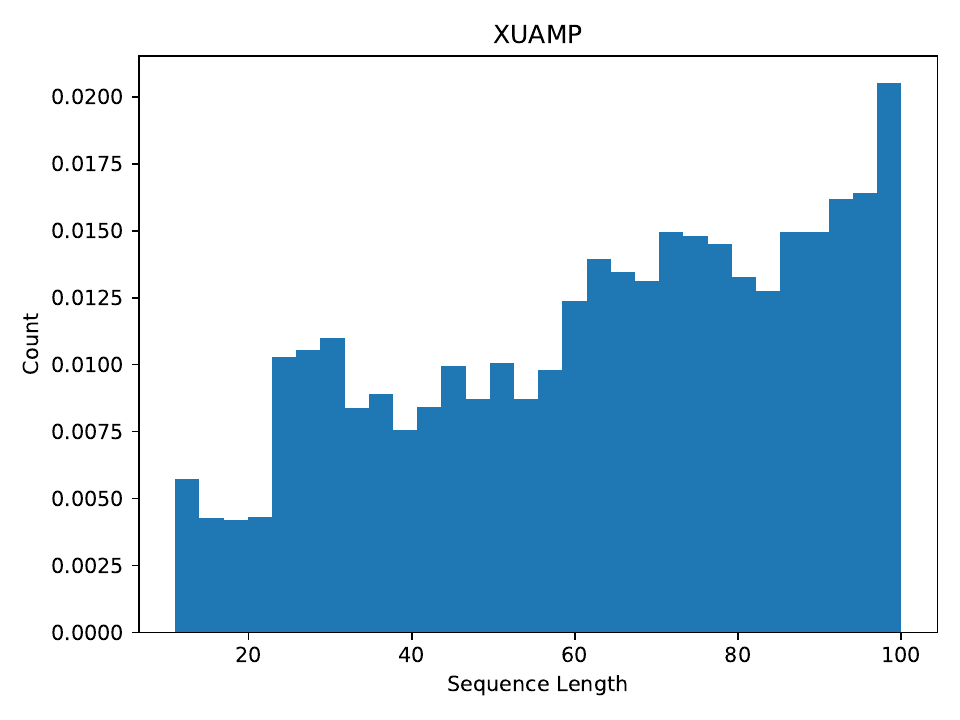}
    \caption{XUAMP \cite{Xu_AMP} sequence length distribution.}
    \label{fig:xuamp_sequence_length_distribution}
\end{figure}

Authors of XUAMP \cite{Xu_AMP} focus solely on antimicrobial peptides. The training part of this one dataset consists of 11072 samples, and testing part of 3072 samples. Both of these subsets have balanced class labels, with 50\% positive and negative samples. We report the sequence length distribution from this dataset in Figure \ref{fig:xuamp_sequence_length_distribution}.

\clearpage

\subsection{LRGB}

Long-Range Graph Benchmark (LRGB) \cite{LRGB} provides multiple datasets, of which we use Peptides-func and Peptides-struct in this work. They have the same peptides, but different labels (classification vs regression). Both datasets contain the same 13204 samples in the training and 2331 in the test set.

\begin{figure}[th!]
    \centering
    \includegraphics[width=0.8\textwidth]{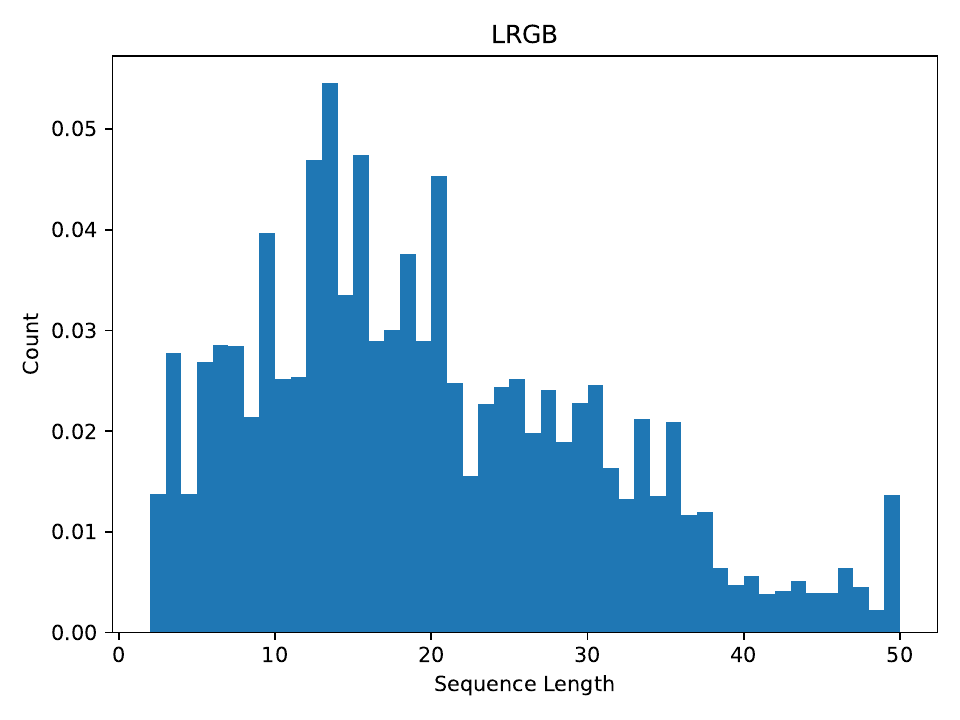}
    \caption{LRGB peptides \cite{LRGB} heavy atom count distribution.}
    \label{fig:lrgb_sequence_length_distributionn}
\end{figure}

We note that LRGB provides valid SMILES for all peptides, but some sequences contain custom notation for chemically modified amino acids. This cannot be parsed by, e.g., ESM or amino acid counts baselines, and also cannot be used for shuffling experiments. Thus, we exclude those sequences for those cases. However, only 125 training sequences and 20 test sequences are removed this way. Based on the remaining, valid sequences we report distribution of sequence lengths as shown in Figure \ref{fig:lrgb_sequence_length_distributionn}

For peptides-func the tasks include antifungal activity, cell–cell communication, anticancer activity, drug delivery vehicle functionality, antimicrobial activity, antiviral activity, antihypertensive activity, antibacterial activity, antiparasitic activity, and toxicity prediction.

For peptides-struct the targets include mass-weighted moments of inertia along principal components 1, 2, and 3, valence-weighted moments of inertia along principal components 1, 2, and 3, peptide length along the x-, y-, and z-axes, sphericity, and deviation from the best-fit plane.

\clearpage

\section{Additional benchmark metrics}
\label{appendix_additional_metrics}

Here, we report additional metrics for AMPs benchmarks from \cite{BERT_benchmark} (Table \ref{appendix_table_bert_benchmark_cd_hit2d}) and \cite{Xu_AMP} (Table \ref{appendix_table_xuamp_metrics}), as well as AutoPeptideML benchmark \cite{AutoPeptideML} (Table \ref{appendix_autopeptideml_metrics}). They were omitted in the main text due to space limitations.

\begin{table}[th!]
\centering
\caption{Results for BERT benchmark \cite{BERT_benchmark} results.}
\resizebox{\textwidth}{!}{
\begin{tabular}{llrrrrrrr}
\hline
\multicolumn{1}{|l|}{\textbf{Metric}}                     & \multicolumn{1}{l|}{\textbf{Model}} & \multicolumn{1}{l|}{\textbf{ADAPTABLE}} & \multicolumn{1}{l|}{\textbf{APD}}    & \multicolumn{1}{l|}{\textbf{CAMP}}  & \multicolumn{1}{l|}{\textbf{dbAMP}} & \multicolumn{1}{l|}{\textbf{DRAMP}} & \multicolumn{1}{l|}{\textbf{YADAMP}} & \multicolumn{1}{l|}{\textbf{avg}}   \\ \hline \hline
\multicolumn{1}{|l|}{\multirow{8}{*}{\textbf{Recall}}}    & \multicolumn{1}{l|}{AMP-BERT}       & \multicolumn{1}{r|}{67.60}              & \multicolumn{1}{r|}{79.40}           & \multicolumn{1}{r|}{80.40}          & \multicolumn{1}{r|}{54.20}          & \multicolumn{1}{r|}{54.70}          & \multicolumn{1}{r|}{82.60}           & \multicolumn{1}{r|}{69.82}          \\ \cline{2-9} 
\multicolumn{1}{|l|}{}                                    & \multicolumn{1}{l|}{Bert-Protein}   & \multicolumn{1}{r|}{58.40}              & \multicolumn{1}{r|}{94.10}           & \multicolumn{1}{r|}{76.30}          & \multicolumn{1}{r|}{83.80}          & \multicolumn{1}{r|}{47.60}          & \multicolumn{1}{r|}{47.60}           & \multicolumn{1}{r|}{67.97}          \\ \cline{2-9} 
\multicolumn{1}{|l|}{}                                    & \multicolumn{1}{l|}{cAMPs\_pred}    & \multicolumn{1}{r|}{44.50}              & \multicolumn{1}{r|}{58.80}           & \multicolumn{1}{r|}{57.90}          & \multicolumn{1}{r|}{36.30}          & \multicolumn{1}{r|}{24.70}          & \multicolumn{1}{r|}{73.90}           & \multicolumn{1}{r|}{49.35}          \\ \cline{2-9} 
\multicolumn{1}{|l|}{}                                    & \multicolumn{1}{l|}{LM\_pred}       & \multicolumn{1}{r|}{40.00}              & \multicolumn{1}{r|}{55.90}           & \multicolumn{1}{r|}{49.10}          & \multicolumn{1}{r|}{45.10}          & \multicolumn{1}{r|}{31.80}          & \multicolumn{1}{r|}{69.60}           & \multicolumn{1}{r|}{48.58}          \\ \cline{2-9} 
\multicolumn{1}{|l|}{}                                    & \multicolumn{1}{l|}{LM\_pred (BFD)} & \multicolumn{1}{r|}{64.60}              & \multicolumn{1}{r|}{88.20}           & \multicolumn{1}{r|}{78.00}          & \multicolumn{1}{r|}{58.50}          & \multicolumn{1}{r|}{52.80}          & \multicolumn{1}{r|}{82.60}           & \multicolumn{1}{r|}{59.67}          \\ \cline{2-9} 
\multicolumn{1}{|l|}{}                                    & \multicolumn{1}{l|}{ECFP}           & \multicolumn{1}{r|}{75.60}              & \multicolumn{1}{r|}{\textbf{97.10}}  & \multicolumn{1}{r|}{88.70}          & \multicolumn{1}{r|}{\textbf{95.10}} & \multicolumn{1}{r|}{66.80}          & \multicolumn{1}{r|}{\textbf{100.00}} & \multicolumn{1}{r|}{87.22}          \\ \cline{2-9} 
\multicolumn{1}{|l|}{}                                    & \multicolumn{1}{l|}{TT}             & \multicolumn{1}{r|}{75.60}              & \multicolumn{1}{r|}{\textbf{97.10}}  & \multicolumn{1}{r|}{\textbf{89.10}} & \multicolumn{1}{r|}{94.00}          & \multicolumn{1}{r|}{\textbf{67.10}} & \multicolumn{1}{r|}{\textbf{100.00}} & \multicolumn{1}{r|}{87.15}          \\ \cline{2-9} 
\multicolumn{1}{|l|}{}                                    & \multicolumn{1}{l|}{RDKit}          & \multicolumn{1}{r|}{\textbf{77.20}}     & \multicolumn{1}{r|}{\textbf{97.10}}  & \multicolumn{1}{r|}{88.70}          & \multicolumn{1}{r|}{94.00}          & \multicolumn{1}{r|}{66.50}          & \multicolumn{1}{r|}{\textbf{100.00}} & \multicolumn{1}{r|}{\textbf{87.25}} \\ \hline \hline
\multicolumn{1}{|l|}{\multirow{8}{*}{\textbf{Precision}}} & \multicolumn{1}{l|}{AMP-BERT}       & \multicolumn{1}{r|}{80.20}              & \multicolumn{1}{r|}{77.10}           & \multicolumn{1}{r|}{84.20}          & \multicolumn{1}{r|}{74.40}          & \multicolumn{1}{r|}{77.10}          & \multicolumn{1}{r|}{82.60}           & \multicolumn{1}{r|}{79.27}          \\ \cline{2-9} 
\multicolumn{1}{|l|}{}                                    & \multicolumn{1}{l|}{Bert-Protein}   & \multicolumn{1}{r|}{79.20}              & \multicolumn{1}{r|}{86.50}           & \multicolumn{1}{r|}{85.40}          & \multicolumn{1}{r|}{82.10}          & \multicolumn{1}{r|}{77.10}          & \multicolumn{1}{r|}{77.10}           & \multicolumn{1}{r|}{81.23}          \\ \cline{2-9} 
\multicolumn{1}{|l|}{}                                    & \multicolumn{1}{l|}{cAMPs\_pred}    & \multicolumn{1}{r|}{\textbf{98.20}}     & \multicolumn{1}{r|}{\textbf{100.00}} & \multicolumn{1}{r|}{\textbf{97.90}} & \multicolumn{1}{r|}{\textbf{94.50}} & \multicolumn{1}{r|}{\textbf{94.80}} & \multicolumn{1}{r|}{\textbf{100.00}} & \multicolumn{1}{r|}{\textbf{97.57}} \\ \cline{2-9} 
\multicolumn{1}{|l|}{}                                    & \multicolumn{1}{l|}{LM\_pred}       & \multicolumn{1}{r|}{92.50}              & \multicolumn{1}{r|}{\textbf{100.00}} & \multicolumn{1}{r|}{94.00}          & \multicolumn{1}{r|}{92.80}          & \multicolumn{1}{r|}{94.00}          & \multicolumn{1}{r|}{\textbf{100.00}} & \multicolumn{1}{r|}{95.55}          \\ \cline{2-9} 
\multicolumn{1}{|l|}{}                                    & \multicolumn{1}{l|}{LM\_pred (BFD)} & \multicolumn{1}{r|}{86.30}              & \multicolumn{1}{r|}{91.20}           & \multicolumn{1}{r|}{90.20}          & \multicolumn{1}{r|}{83.40}          & \multicolumn{1}{r|}{83.90}          & \multicolumn{1}{r|}{86.40}           & \multicolumn{1}{r|}{91.62}          \\ \cline{2-9} 
\multicolumn{1}{|l|}{}                                    & \multicolumn{1}{l|}{ECFP}           & \multicolumn{1}{r|}{86.90}              & \multicolumn{1}{r|}{89.20}           & \multicolumn{1}{r|}{87.70}          & \multicolumn{1}{r|}{88.20}          & \multicolumn{1}{r|}{89.00}          & \multicolumn{1}{r|}{92.00}           & \multicolumn{1}{r|}{88.83}          \\ \cline{2-9} 
\multicolumn{1}{|l|}{}                                    & \multicolumn{1}{l|}{TT}             & \multicolumn{1}{r|}{86.40}              & \multicolumn{1}{r|}{91.70}           & \multicolumn{1}{r|}{88.50}          & \multicolumn{1}{r|}{87.80}          & \multicolumn{1}{r|}{89.80}          & \multicolumn{1}{r|}{92.00}           & \multicolumn{1}{r|}{89.37}          \\ \cline{2-9} 
\multicolumn{1}{|l|}{}                                    & \multicolumn{1}{l|}{RDKit}          & \multicolumn{1}{r|}{87.50}              & \multicolumn{1}{r|}{89.20}           & \multicolumn{1}{r|}{88.00}          & \multicolumn{1}{r|}{89.00}          & \multicolumn{1}{r|}{89.80}          & \multicolumn{1}{r|}{85.20}           & \multicolumn{1}{r|}{88.12}          \\ \hline \hline
\multicolumn{1}{|l|}{\multirow{8}{*}{\textbf{AUROC}}}     & \multicolumn{1}{l|}{AMP-BERT}       & \multicolumn{1}{r|}{81.70}              & \multicolumn{1}{r|}{91.80}           & \multicolumn{1}{r|}{90.60}          & \multicolumn{1}{r|}{78.40}          & \multicolumn{1}{r|}{76.90}          & \multicolumn{1}{r|}{90.70}           & \multicolumn{1}{r|}{85.02}          \\ \cline{2-9} 
\multicolumn{1}{|l|}{}                                    & \multicolumn{1}{l|}{Bert-Protein}   & \multicolumn{1}{r|}{79.40}              & \multicolumn{1}{r|}{96.10}           & \multicolumn{1}{r|}{90.90}          & \multicolumn{1}{r|}{91.90}          & \multicolumn{1}{r|}{74.00}          & \multicolumn{1}{r|}{74.00}           & \multicolumn{1}{r|}{84.38}          \\ \cline{2-9} 
\multicolumn{1}{|l|}{}                                    & \multicolumn{1}{l|}{cAMPs\_pred}    & \multicolumn{1}{r|}{82.30}              & \multicolumn{1}{r|}{91.80}           & \multicolumn{1}{r|}{87.00}          & \multicolumn{1}{r|}{81.90}          & \multicolumn{1}{r|}{68.10}          & \multicolumn{1}{r|}{93.00}           & \multicolumn{1}{r|}{84.02}          \\ \cline{2-9} 
\multicolumn{1}{|l|}{}                                    & \multicolumn{1}{l|}{LM\_pred}       & \multicolumn{1}{r|}{88.00}              & \multicolumn{1}{r|}{97.20}           & \multicolumn{1}{r|}{92.30}          & \multicolumn{1}{r|}{89.50}          & \multicolumn{1}{r|}{82.80}          & \multicolumn{1}{r|}{98.70}           & \multicolumn{1}{r|}{91.42}          \\ \cline{2-9} 
\multicolumn{1}{|l|}{}                                    & \multicolumn{1}{l|}{LM\_pred (BFD)} & \multicolumn{1}{r|}{86.30}              & \multicolumn{1}{r|}{97.40}           & \multicolumn{1}{r|}{93.60}          & \multicolumn{1}{r|}{87.60}          & \multicolumn{1}{r|}{83.90}          & \multicolumn{1}{r|}{93.40}           & \multicolumn{1}{r|}{90.47}          \\ \cline{2-9} 
\multicolumn{1}{|l|}{}                                    & \multicolumn{1}{l|}{ECFP}           & \multicolumn{1}{r|}{89.50}              & \multicolumn{1}{r|}{98.40}           & \multicolumn{1}{r|}{94.30}          & \multicolumn{1}{r|}{\textbf{97.20}} & \multicolumn{1}{r|}{87.80}          & \multicolumn{1}{r|}{\textbf{100.00}} & \multicolumn{1}{r|}{94.53}          \\ \cline{2-9} 
\multicolumn{1}{|l|}{}                                    & \multicolumn{1}{l|}{TT}             & \multicolumn{1}{r|}{89.30}              & \multicolumn{1}{r|}{98.20}           & \multicolumn{1}{r|}{94.30}          & \multicolumn{1}{r|}{97.10}          & \multicolumn{1}{r|}{87.60}          & \multicolumn{1}{r|}{\textbf{100.00}} & \multicolumn{1}{r|}{94.42}          \\ \cline{2-9} 
\multicolumn{1}{|l|}{}                                    & \multicolumn{1}{l|}{RDKit}          & \multicolumn{1}{r|}{\textbf{90.20}}     & \multicolumn{1}{r|}{\textbf{99.30}}  & \multicolumn{1}{r|}{\textbf{94.80}} & \multicolumn{1}{r|}{97.00}          & \multicolumn{1}{r|}{\textbf{88.20}} & \multicolumn{1}{r|}{\textbf{100.00}} & \multicolumn{1}{r|}{\textbf{94.92}} \\ \hline \hline
\multicolumn{1}{|l|}{\multirow{8}{*}{\textbf{F1}}}        & \multicolumn{1}{l|}{AMP-BERT}       & \multicolumn{1}{r|}{81.70}              & \multicolumn{1}{r|}{78.30}           & \multicolumn{1}{r|}{82.30}          & \multicolumn{1}{r|}{62.70}          & \multicolumn{1}{r|}{64.00}          & \multicolumn{1}{r|}{82.60}           & \multicolumn{1}{r|}{75.27}          \\ \cline{2-9} 
\multicolumn{1}{|l|}{}                                    & \multicolumn{1}{l|}{Bert-Protein}   & \multicolumn{1}{r|}{79.40}              & \multicolumn{1}{r|}{90.10}           & \multicolumn{1}{r|}{80.60}          & \multicolumn{1}{r|}{82.90}          & \multicolumn{1}{r|}{58.80}          & \multicolumn{1}{r|}{58.80}           & \multicolumn{1}{r|}{75.10}          \\ \cline{2-9} 
\multicolumn{1}{|l|}{}                                    & \multicolumn{1}{l|}{cAMPs\_pred}    & \multicolumn{1}{r|}{61.30}              & \multicolumn{1}{r|}{74.10}           & \multicolumn{1}{r|}{72.80}          & \multicolumn{1}{r|}{52.40}          & \multicolumn{1}{r|}{39.20}          & \multicolumn{1}{r|}{85.00}           & \multicolumn{1}{r|}{64.13}          \\ \cline{2-9} 
\multicolumn{1}{|l|}{}                                    & \multicolumn{1}{l|}{LM\_pred}       & \multicolumn{1}{r|}{55.90}              & \multicolumn{1}{r|}{71.70}           & \multicolumn{1}{r|}{64.50}          & \multicolumn{1}{r|}{60.70}          & \multicolumn{1}{r|}{47.60}          & \multicolumn{1}{r|}{82.10}           & \multicolumn{1}{r|}{63.75}          \\ \cline{2-9} 
\multicolumn{1}{|l|}{}                                    & \multicolumn{1}{l|}{LM\_pred (BFD)} & \multicolumn{1}{r|}{73.90}              & \multicolumn{1}{r|}{89.60}           & \multicolumn{1}{r|}{83.60}          & \multicolumn{1}{r|}{68.70}          & \multicolumn{1}{r|}{64.80}          & \multicolumn{1}{r|}{84.40}           & \multicolumn{1}{r|}{70.32}          \\ \cline{2-9} 
\multicolumn{1}{|l|}{}                                    & \multicolumn{1}{l|}{ECFP}           & \multicolumn{1}{r|}{80.80}              & \multicolumn{1}{r|}{93.00}           & \multicolumn{1}{r|}{88.20}          & \multicolumn{1}{r|}{\textbf{91.50}} & \multicolumn{1}{r|}{76.30}          & \multicolumn{1}{r|}{\textbf{95.80}}  & \multicolumn{1}{r|}{87.60}          \\ \cline{2-9} 
\multicolumn{1}{|l|}{}                                    & \multicolumn{1}{l|}{TT}             & \multicolumn{1}{r|}{80.60}              & \multicolumn{1}{r|}{\textbf{94.30}}  & \multicolumn{1}{r|}{\textbf{88.80}} & \multicolumn{1}{r|}{90.80}          & \multicolumn{1}{r|}{\textbf{76.90}} & \multicolumn{1}{r|}{\textbf{95.80}}  & \multicolumn{1}{r|}{\textbf{87.87}} \\ \cline{2-9} 
\multicolumn{1}{|l|}{}                                    & \multicolumn{1}{l|}{RDKit}          & \multicolumn{1}{r|}{\textbf{82.00}}     & \multicolumn{1}{r|}{93.00}           & \multicolumn{1}{r|}{88.40}          & \multicolumn{1}{r|}{91.40}          & \multicolumn{1}{r|}{76.40}          & \multicolumn{1}{r|}{92.00}           & \multicolumn{1}{r|}{87.20}          \\ \hline
\end{tabular}
}
\label{appendix_table_bert_benchmark_cd_hit2d}
\end{table}

\clearpage

\begin{table}[th!]
\centering
\caption{Additional metrics for results on \cite{Xu_AMP}.}
\begin{tabular}{|c|c|c|c|c|c|c|}
\hline
\textbf{Method} & \textbf{Accuracy} & \textbf{AUROC} & \textbf{F1}   & \textbf{MCC}   & \textbf{Recall} & \textbf{Specificity} \\ \hline
AMPscannerV2    & 56.8              & 58.5           & 54.8          & 0.137          & \textbf{52.3}   & 61.3                 \\ \hline
iAMP-2L         & 59.2              & 59.2           & 36.8          & 0.261          & 23.8            & 94.7                 \\ \hline
ADAM-SVM        & 61.2              & 61.2           & 47.1          & 0.264          & 34.6            & 87.8                 \\ \hline
ampir           & 56.3              & 61.9           & 37.9          & 0.156          & 26.6            & 85.9                 \\ \hline
MLAMP           & 55.3              & 62.9           & 23            & 0.194          & 13.3            & 97.2                 \\ \hline
ADAM-HMM        & 68.4              & 68.4           & \textbf{62.3} & 0.39           & 52.1            & 84.7                 \\ \hline
AMPlify         & 64.2              & 69.7           & 46.2          & 0.381          & 30.7            & 97.6                 \\ \hline
AMPEP           & 65.8              & 72.7           & 48.7          & 0.425          & 32.5            & \textbf{99.2}        \\ \hline
AMPfun          & 67.4              & 73.5           & 55.5          & 0.414          & 40.6            & 94.3                 \\ \hline
\hline
RDKit           & 67.5              & 73.7           & 58.5          & 0.388          & 45.8            & 89.1                 \\ \hline
ECFP            & 69.3              & 75.3           & 60.9          & \textbf{0.426} & 47.9            & 90.6                 \\ \hline
TT              & \textbf{69.4}     & \textbf{74.9}  & 61.7          & 0.424          & 49.3            & 89.5                 \\ \hline
\end{tabular}
\label{appendix_table_xuamp_metrics}
\end{table}

\begin{table}[th!]
\centering
\caption{Additional metrics for results on AutoPeptideML \cite{AutoPeptideML}.}
\begin{tabular}{llllll}
\hline
\multicolumn{1}{|l|}{\textbf{Model}} & \multicolumn{1}{l|}{\textbf{\# params}} & \multicolumn{1}{l|}{\textbf{Accuracy}} & \multicolumn{1}{l|}{\textbf{MCC}}   & \multicolumn{1}{l|}{\textbf{AUROC}} & \multicolumn{1}{l|}{\textbf{F1}}    \\ \hline
\multicolumn{1}{|l|}{ProtBERT}       & \multicolumn{1}{l|}{420M}               & \multicolumn{1}{l|}{68.5}             & \multicolumn{1}{l|}{0.375}          & \multicolumn{1}{l|}{75.9}          & \multicolumn{1}{l|}{67.7}          \\ \hline
\multicolumn{1}{|l|}{ESM2-150M}      & \multicolumn{1}{l|}{150M}               & \multicolumn{1}{l|}{69.8}             & \multicolumn{1}{l|}{0.402}          & \multicolumn{1}{l|}{77.7}          & \multicolumn{1}{l|}{68.5}          \\ \hline
\multicolumn{1}{|l|}{Prost-T5}       & \multicolumn{1}{l|}{3B}                 & \multicolumn{1}{l|}{70.0}             & \multicolumn{1}{l|}{0.409}          & \multicolumn{1}{l|}{77.1}          & \multicolumn{1}{l|}{69.0}          \\ \hline
\multicolumn{1}{|l|}{ESM2-8M}        & \multicolumn{1}{l|}{8M}                 & \multicolumn{1}{l|}{70.2}             & \multicolumn{1}{l|}{0.418}          & \multicolumn{1}{l|}{77.5}          & \multicolumn{1}{l|}{69.4}          \\ \hline
\multicolumn{1}{|l|}{ESM2-35M}       & \multicolumn{1}{l|}{35M}                & \multicolumn{1}{l|}{71.0}             & \multicolumn{1}{l|}{0.428}          & \multicolumn{1}{l|}{78.0}          & \multicolumn{1}{l|}{70.2}          \\ \hline
\multicolumn{1}{|l|}{ESM1b-650M}     & \multicolumn{1}{l|}{650M}               & \multicolumn{1}{l|}{71.1}             & \multicolumn{1}{l|}{0.433}          & \multicolumn{1}{l|}{78.9}          & \multicolumn{1}{l|}{69.2}          \\ \hline
\multicolumn{1}{|l|}{ESM2-650M}      & \multicolumn{1}{l|}{650M}               & \multicolumn{1}{l|}{68.0}             & \multicolumn{1}{l|}{0.438}          & \multicolumn{1}{l|}{\textbf{79.7}} & \multicolumn{1}{l|}{69.7}          \\ \hline
\multicolumn{1}{|l|}{Prot-T5-XL}     & \multicolumn{1}{l|}{3B}                 & \multicolumn{1}{l|}{68.9}             & \multicolumn{1}{l|}{\textbf{0.447}} & \multicolumn{1}{l|}{\textbf{79.7}} & \multicolumn{1}{l|}{\textbf{70.4}} \\ \hline \hline
\multicolumn{1}{|l|}{RDKit}          & \multicolumn{1}{l|}{\textbf{20k}}                & \multicolumn{1}{l|}{70.7}             & \multicolumn{1}{l|}{0.421}          & \multicolumn{1}{l|}{76.9}          & \multicolumn{1}{l|}{69.6}          \\ \hline
\multicolumn{1}{|l|}{TT}             & \multicolumn{1}{l|}{\textbf{23k}}                & \multicolumn{1}{l|}{70.8}             & \multicolumn{1}{l|}{0.422}          & \multicolumn{1}{l|}{77.1}          & \multicolumn{1}{l|}{69.4}          \\ \hline
\multicolumn{1}{|l|}{ECFP}           & \multicolumn{1}{l|}{\textbf{22k}}                & \multicolumn{1}{l|}{\textbf{71.5}}    & \multicolumn{1}{l|}{0.437}          & \multicolumn{1}{l|}{78.1}          & \multicolumn{1}{l|}{70.2}          \\ \hline
\end{tabular}
\label{appendix_autopeptideml_metrics}
\end{table}

\section{Additional classifiers on LRGB}
\label{appendix_additional_lrgb_classifiers}

Here, we present results of two additional classifiers on the LRGB benchmark datasets: Random Forest and Extremely Randomized Trees \cite{extremely_randomized_trees}. For both we use 500 trees, entropy or squared error as cost function (for classification and regression, respectively), and class weighting for classification. We can report standard deviations for them over 10 random seeds, because they are nondeterministic.

Peptides-func results are in Table \ref{appendix_peptides_func_additional_classifiers}, and Peptides-struct in Table \ref{appendix_peptides_struct_additional_classifiers}. For clarity, we also include LightGBM results from the main body, which is a deterministic classifier and thus does not have standard deviation.

Random Forest and Extremely Randomized Trees classifiers have very low standard deviation, highlighting the stability of the fingerprint-based approach. Models based on ECFP features consistently achieve the best results in all cases, but other fingerprints and classifiers also obtain results highly competitive with long-range GNNs.

\begin{table}[h!]
\centering
\caption{AUPRC $\uparrow$ for different classifiers on Peptides-func.}
\begin{tabular}{|c|c|c|c|}
\hline
\textbf{}      & \textbf{LightGBM} & \textbf{Random Forest} & \textbf{Extremely Randomized Trees} \\ \hline
\textbf{RDKit} & 73.11 & 71.48 $\pm$ 0.13       & 69.38 $\pm$ 0.06     \\ \hline
\textbf{TT}    & 73.18 & 71.66 $\pm$ 0.09       & 69.86 $\pm$ 0.08     \\ \hline
\textbf{ECFP}  & 74.60 & 73.55 $\pm$ 0.10       & 71.98 $\pm$ 0.08     \\ \hline
\end{tabular}
\label{appendix_peptides_func_additional_classifiers}
\end{table}

\begin{table}[h!]
\centering
\caption{MAE $\downarrow$ for different classifiers on Peptides-struct.}
\begin{tabular}{|c|c|c|c|}
\hline
\textbf{}      & \textbf{LightGBM} &\textbf{Random Forest} & \textbf{Extremely Randomized Trees} \\ \hline
\textbf{RDKit} & 0.2459 & 0.2459 $\pm$ 0.0002       & 0.2440 $\pm$ 0.0001     \\ \hline
\textbf{TT}    & 0.2438 & 0.2471 $\pm$ 0.0003       & 0.2467 $\pm$ 0.0003     \\ \hline
\textbf{ECFP}  & 0.2432 & 0.2442 $\pm$ 0.0002       & 0.2433 $\pm$ 0.0001     \\ \hline
\end{tabular}
\label{appendix_peptides_struct_additional_classifiers}
\end{table}

\newpage

\section{Additional LRGB timings}
\label{appendix_additional_timings}

Here, we present expanded timings of fingerprints and compare them to times reported in LRGB \cite{LRGB} (Appendix C.2), summarizing them in Table \ref{table_lrgb_timings}. Times for fingerprint-based models are an average of 10 runs on 12-core Intel Core i7-12700KF. We measure them for \textit{Peptides-func}, and for \textit{Peptides-struct} results were almost identical. Timings for GNNs are taken from \cite{LRGB}, and use Nvidia A100 GPU.

The entire time for our approach (feature extraction + classifier training) is shorter than precomputing structural embeddings or training even a single epoch of SAN model. Graph transformer model has faster epochs, they still require precomputing LapPE embeddings, while also giving much worse results than fingerprints (see main body for results table).

One should also take into consideration the difference in raw compute power (CPU vs powerful GPU) required to get those times, which makes the difference even more significant.

\begin{table}[h!]
\centering
\caption{Time of computation on Peptides-func.}
\begin{tabular}{|c|c|}
\hline
\textbf{Model} & \textbf{Time {[}s{]}} \\ \hline
Transformer+LapPE & 5.9 \\ \hline
SAN+LapPE & 53.6 \\ \hline
SAN+RWSE & 49.7 \\ \hline
LapPE encoding & 74 \\ \hline
RWSE encoding & 53 \\ \hline
\hline
ECFP & 19 \\ \hline
TT & 15.2 \\ \hline
RDKit & 42.8 \\ \hline
\end{tabular}
\label{table_lrgb_timings}
\end{table}

\clearpage

\section{Additional PeptideReactor results}
\label{appendix_additional_peptidereactor_results}

Here, we present the performance of all encodings on PeptideReactor \cite{PeptideReactor}, in Table \ref{appendix_table_results_peptidereactor}. They are sorted from highest to lowest result. Further, in Table \ref{appendix_table_peptidereactor_untuned}, we provide results for molecular fingerprints with and without tuning their hyperparameters.

\begin{table}[h!]
\centering
\caption{Full results on PeptideReactor benchmark \cite{PeptideReactor}.}
\resizebox{0.32\textwidth}{!}{
\begin{tabular}{|c|c|c|}
\hline
\textbf{Encoding} & \textbf{Avg F1} & \textbf{Type} \\ \hline
FP encoding       & 82.9            & fingerprint   \\ \hline
cksaap            & 82.4            & sequence      \\ \hline
dist\_f           & 82.2            & sequence      \\ \hline
psekraac          & 81.8            & sequence      \\ \hline
ECFP        & 81.7            & fingerprint   \\ \hline
ngram\_           & 81.6            & sequence      \\ \hline
dde               & 81.3            & sequence      \\ \hline
dpc               & 80.6            & sequence      \\ \hline
TT          & 80.6            & fingerprint   \\ \hline
RDKit       & 80.6            & fingerprint   \\ \hline
fldpc\_           & 79.4            & sequence      \\ \hline
qsorde            & 78.9            & sequence      \\ \hline
waac\_a           & 78.8            & sequence      \\ \hline
apaac\_           & 78.2            & sequence      \\ \hline
aac               & 78.1            & sequence      \\ \hline
binary            & 77.9            & sequence      \\ \hline
ctdd              & 77.9            & sequence      \\ \hline
paac\_l           & 77.9            & sequence      \\ \hline
aainde            & 77.8            & sequence      \\ \hline
tpc               & 76.6            & sequence      \\ \hline
cksaag            & 76.5            & sequence      \\ \hline
ctriad            & 75.9            & sequence      \\ \hline
ctdt              & 75.7            & sequence      \\ \hline
ctdc              & 75.6            & sequence      \\ \hline
ksctri            & 75.6            & sequence      \\ \hline
gtpc              & 75              & sequence      \\ \hline
gdpc              & 74              & sequence      \\ \hline
flgc\_a           & 73.8            & sequence      \\ \hline
fft\_aa           & 73.3            & sequence      \\ \hline
qsar              & 72.8            & structure     \\ \hline
nmbrot            & 72.3            & sequence      \\ \hline
zscale            & 72.1            & sequence      \\ \hline
eaac\_w           & 71.6            & sequence      \\ \hline
blomap            & 70.9            & sequence      \\ \hline
egaac\_           & 70.3            & sequence      \\ \hline
blosum            & 69.8            & sequence      \\ \hline
delaun            & 69.5            & structure     \\ \hline
gaac              & 69.5            & sequence      \\ \hline
moran\_           & 69              & sequence      \\ \hline
socnum            & 69              & sequence      \\ \hline
geary\_           & 68.9            & sequence      \\ \hline
cgr\_re           & 68.8            & sequence      \\ \hline
distan            & 61.3            & structure     \\ \hline
electr            & 60.9            & structure     \\ \hline
disord            & 53.5            & structure     \\ \hline
sseb              & 51.7            & structure     \\ \hline
asa               & 51.5            & structure     \\ \hline
ta                & 51.4            & structure     \\ \hline
ssec              & 48.3            & structure     \\ \hline
\end{tabular}
}
\label{appendix_table_results_peptidereactor}
\end{table}

\begin{table}[]
\centering
\caption{Results on PeptideReactor benchmark \cite{PeptideReactor} with and without tuning.}
\begin{tabular}{|c|c|c|}
\hline
\textbf{Fingerprint} & \textbf{Tuned?} & \textbf{F1 score} \\ \hline
\multirow{2}{*}{ECFP} & No & 79.4\% \\ \cline{2-3} 
 & Yes & 81.7\% \\ \hline
\multirow{2}{*}{\begin{tabular}[c]{@{}c@{}}Topological\\ Torsion\end{tabular}} & No & 77.8\% \\ \cline{2-3} 
 & Yes & 80.6\% \\ \hline
\multirow{2}{*}{RDKit} & No & 80.0\% \\ \cline{2-3} 
 & Yes & 80.6\% \\ \hline
\end{tabular}
\label{appendix_table_peptidereactor_untuned}
\end{table}

\clearpage

\section{Shuffling results}
\label{appendix_shuffling_results}

Here we report additional sequence shuffling experiments. We evaluate the performance of three models for different shuffling ratios: molecular fingerprints ECFP model, amino acid counts, and ESM2. For each benchmark and model we perform two distinct experiments. In one, we shuffle only sequences from the train set, and in the other from both train and test set.

\begin{figure}[th!]
    \centering
    \includegraphics[width=0.8\textwidth]{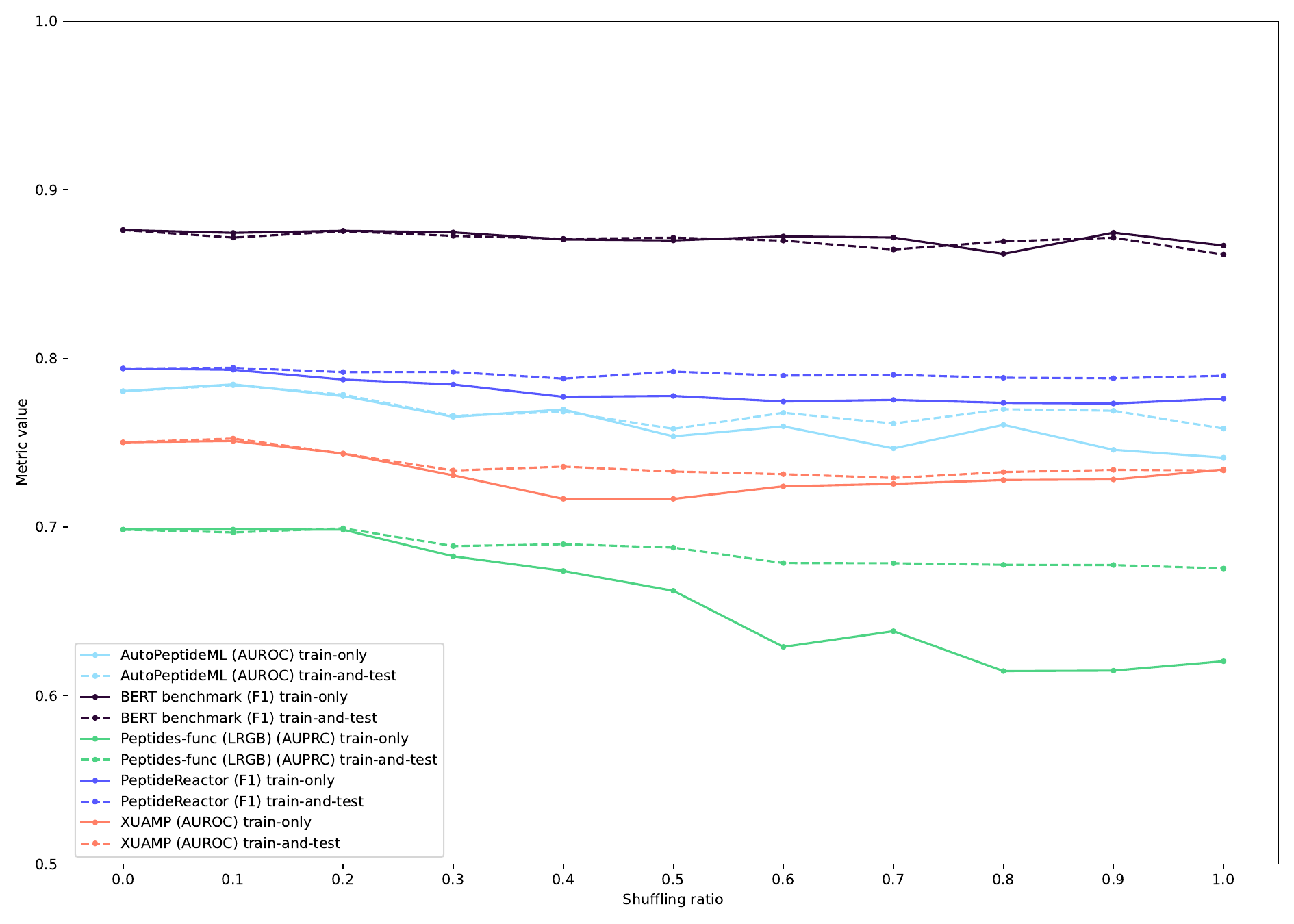}
    \caption{Shuffling  metrics for ECFP + LightGBM.}
    \label{fig:shuffling_metrics_ECFP}
\end{figure}

\begin{figure}[th!]
    \centering
    \includegraphics[width=0.8\textwidth]{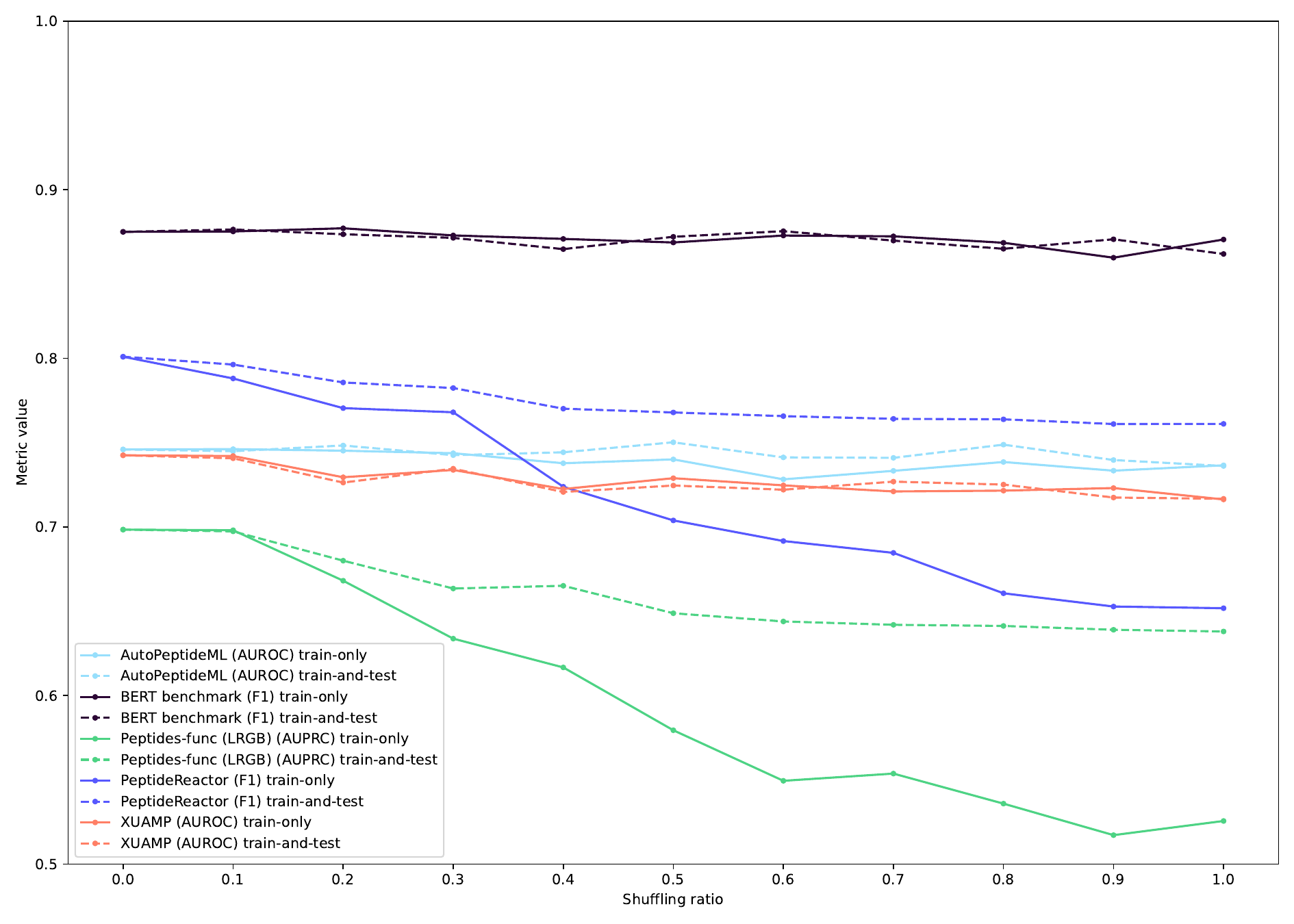}
    \caption{Shuffling  metrics for amino acid counts model.}
    \label{fig:shuffling_metrics_aminoacid_counts}
\end{figure}

\begin{figure}[th!]
    \centering
    \includegraphics[width=0.8\textwidth]{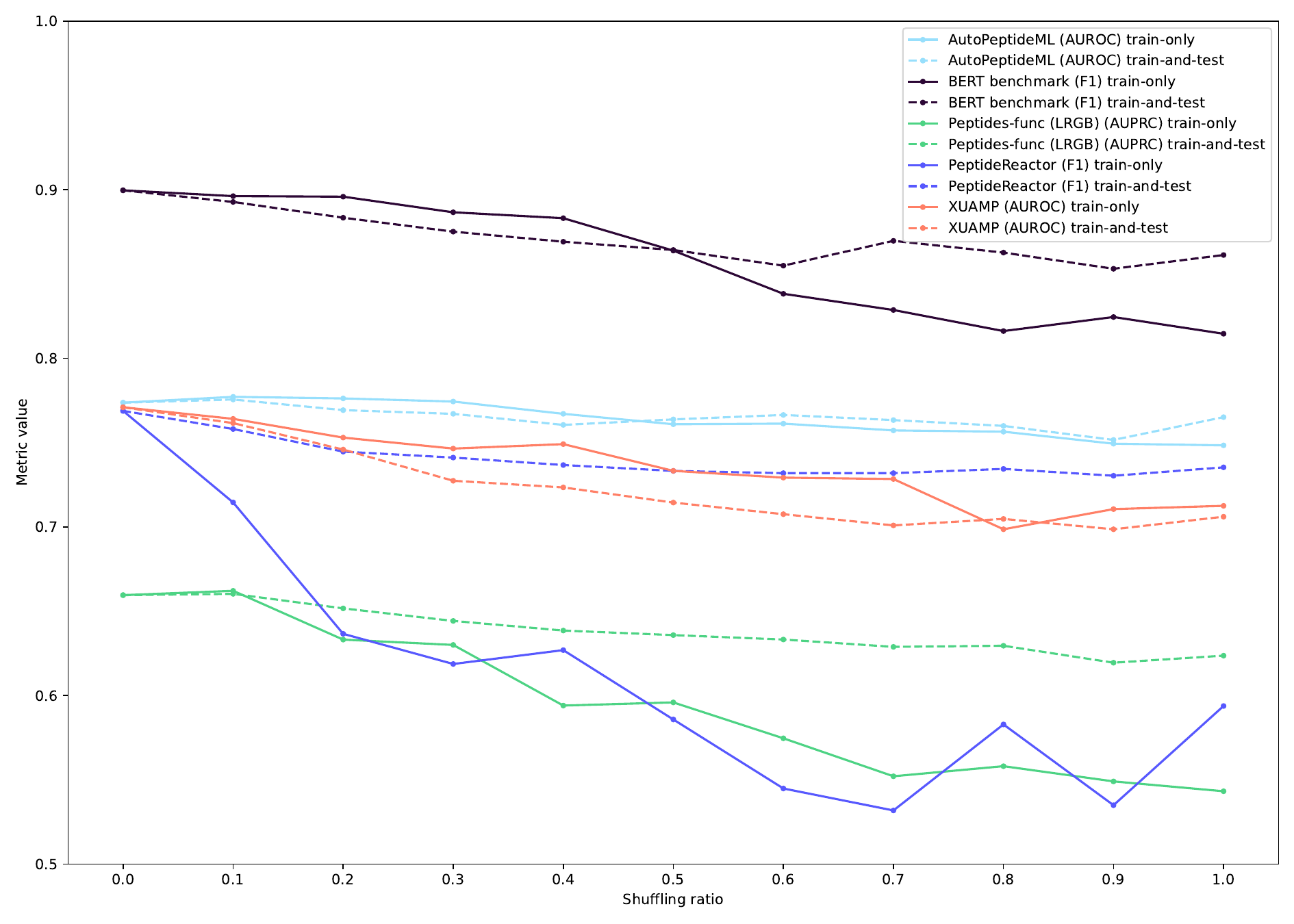}
    \caption{Shuffling  metrics for ESM2 \cite{ESM2}.}
    \label{fig:shuffling_metrics_ESM}
\end{figure}

We present results for individual models in Figures \ref{fig:shuffling_metrics_ECFP}, \ref{fig:shuffling_metrics_aminoacid_counts} and \ref{fig:shuffling_metrics_ESM}. Comparisons of those models within each benchmark are reported in Tables \ref{autopeptideml_shuffling_results}, \ref{bert_amp_shuffling_results}, \ref{lrgb_shuffling_results}, \ref{peptidereactor_shuffling_results} and \ref{xuamp_shuffling_results}. Results include the performance loss delta, computed as the difference between performance achieved with unshuffled and fully shuffled data. 

For both variants of the experiment, our ECFP model maintains comparable performance regardless of shuffling ratio, showing only a slight decrease compared to other models. Amino acid count model performs consistently only on AutoPeptideML, BERT AMP and XUAMP. Degradation in performance of sequence-based baselines is particularly prominent for LRGB and PeptideReactor. We can see that the two baseline models perform worse in the unshuffled test set experiment. Our hypothesis is that this induces a distribution shift between shuffled train and unshuffled test sequences, resulting in quality degradation.

\begin{table}[h!]
\centering
\caption{Shuffling results for AutoPeptideML \cite{AutoPeptideML}, AUROC metric.}
\begin{tabular}{|c|cc|cc|cc|}
\hline
\multicolumn{1}{|c|}{\multirow{2}{*}{\textbf{Shuffle ratio}}} & \multicolumn{2}{c|}{\textbf{ECFP}}                      & \multicolumn{2}{c|}{\textbf{Amino acid counts}}          & \multicolumn{2}{c|}{\textbf{ESM2}}                       \\ \cline{2-7} 
\multicolumn{1}{|c|}{}                                      & \multicolumn{1}{l|}{\textbf{train}} & \textbf{train and test} & \multicolumn{1}{l|}{\textbf{train}} & \textbf{train and test} & \multicolumn{1}{l|}{\textbf{train}} & \textbf{train and test} \\ \hline
0.0                                                         & \multicolumn{1}{l|}{78.05}          & 78.05                   & \multicolumn{1}{l|}{74.59}          & 74.59                   & \multicolumn{1}{l|}{77.37}          & 77.37                   \\ \hline
0.1                                                         & \multicolumn{1}{l|}{78.45}          & 78.4                    & \multicolumn{1}{l|}{74.61}          & 74.49                   & \multicolumn{1}{l|}{77.71}          & 77.56                   \\ \hline
0.2                                                         & \multicolumn{1}{l|}{77.76}          & 77.85                   & \multicolumn{1}{l|}{74.52}          & 74.83                   & \multicolumn{1}{l|}{77.62}          & 76.93                   \\ \hline
0.3                                                         & \multicolumn{1}{l|}{76.53}          & 76.59                   & \multicolumn{1}{l|}{74.37}          & 74.26                   & \multicolumn{1}{l|}{77.44}          & 76.71                   \\ \hline
0.4                                                         & \multicolumn{1}{l|}{76.97}          & 76.84                   & \multicolumn{1}{l|}{73.78}          & 74.43                   & \multicolumn{1}{l|}{76.71}          & 76.05                   \\ \hline
0.5                                                         & \multicolumn{1}{l|}{75.37}          & 75.82                   & \multicolumn{1}{l|}{74.00}             & 75.02                   & \multicolumn{1}{l|}{76.09}          & 76.38                   \\ \hline
0.6                                                         & \multicolumn{1}{l|}{75.96}          & 76.77                   & \multicolumn{1}{l|}{72.82}          & 74.12                   & \multicolumn{1}{l|}{76.13}          & 76.64                   \\ \hline
0.7                                                         & \multicolumn{1}{l|}{74.66}          & 76.13                   & \multicolumn{1}{l|}{73.33}          & 74.10                    & \multicolumn{1}{l|}{75.72}          & 76.34                   \\ \hline
0.8                                                         & \multicolumn{1}{l|}{76.05}          & 76.98                   & \multicolumn{1}{l|}{73.85}          & 74.88                   & \multicolumn{1}{l|}{75.65}          & 75.99                   \\ \hline
0.9                                                         & \multicolumn{1}{l|}{74.57}          & 76.89                   & \multicolumn{1}{l|}{73.34}          & 73.97                   & \multicolumn{1}{l|}{74.94}          & 75.16                   \\ \hline
1.0                                                         & \multicolumn{1}{l|}{74.11}          & 75.83                   & \multicolumn{1}{l|}{73.65}          & 73.61                   & \multicolumn{1}{l|}{74.84}          & 76.51                   \\ \hline
\hline
delta                                                       & \multicolumn{1}{l|}{-3.94}          & -2.22                   & \multicolumn{1}{l|}{-0.94}          & -0.98                   & \multicolumn{1}{l|}{-2.53}          & -0.86                   \\ \hline
\end{tabular}
\label{autopeptideml_shuffling_results}
\end{table}

\begin{table}[h!]
\centering
\caption{Shuffling results for BERT-based models benchmark \cite{BERT_benchmark}, F1 metric.}
\begin{tabular}{|c|cc|cc|cc|}
\hline
\multirow{2}{*}{\textbf{Shuffle ratio}} & \multicolumn{2}{c|}{\textbf{ECFP}} & \multicolumn{2}{c|}{\textbf{Amino acid counts}} & \multicolumn{2}{c|}{\textbf{ESM2}} \\ \cline{2-7} 
 & \multicolumn{1}{c|}{\textbf{train}} & \textbf{train and test} & \multicolumn{1}{c|}{\textbf{train}} & \textbf{train and test} & \multicolumn{1}{c|}{\textbf{train}} & \textbf{train and test} \\ \hline
0.0 & \multicolumn{1}{c|}{87.61} & 87.61 & \multicolumn{1}{c|}{87.5} & 87.5 & \multicolumn{1}{c|}{89.96} & 89.96 \\ \hline
0.1 & \multicolumn{1}{c|}{87.44} & 87.16 & \multicolumn{1}{c|}{87.52} & 87.64 & \multicolumn{1}{c|}{89.62} & 89.27 \\ \hline
0.2 & \multicolumn{1}{c|}{87.56} & 87.53 & \multicolumn{1}{c|}{87.71} & 87.36 & \multicolumn{1}{c|}{89.58} & 88.34 \\ \hline
0.3 & \multicolumn{1}{c|}{87.47} & 87.26 & \multicolumn{1}{c|}{87.29} & 87.14 & \multicolumn{1}{c|}{88.66} & 87.51 \\ \hline
0.4 & \multicolumn{1}{c|}{87.05} & 87.09 & \multicolumn{1}{c|}{87.08} & 86.47 & \multicolumn{1}{c|}{88.31} & 86.91 \\ \hline
0.5 & \multicolumn{1}{c|}{86.99} & 87.15 & \multicolumn{1}{c|}{86.87} & 87.21 & \multicolumn{1}{c|}{86.39} & 86.42 \\ \hline
0.6 & \multicolumn{1}{c|}{87.23} & 86.98 & \multicolumn{1}{c|}{87.28} & 87.54 & \multicolumn{1}{c|}{83.83} & 85.49 \\ \hline
0.7 & \multicolumn{1}{c|}{87.16} & 86.45 & \multicolumn{1}{c|}{87.23} & 86.98 & \multicolumn{1}{c|}{82.86} & 86.97 \\ \hline
0.8 & \multicolumn{1}{c|}{86.20} & 86.93 & \multicolumn{1}{c|}{86.85} & 86.49 & \multicolumn{1}{c|}{81.62} & 86.27 \\ \hline
0.9 & \multicolumn{1}{c|}{87.45} & 87.15 & \multicolumn{1}{c|}{85.97} & 87.06 & \multicolumn{1}{c|}{82.45} & 85.31 \\ \hline
1.0 & \multicolumn{1}{c|}{86.68} & 86.16 & \multicolumn{1}{c|}{87.04} & 86.19 & \multicolumn{1}{c|}{81.46} & 86.12 \\ \hline
\hline
delta & \multicolumn{1}{c|}{-0.93} & -1.45 & \multicolumn{1}{c|}{-0.46} & -1.31 & \multicolumn{1}{c|}{-8.50} & -3.84 \\ \hline
\end{tabular}
\label{bert_amp_shuffling_results}
\end{table}

\begin{table}[h!]
\centering
\caption{Shuffling results for LRGB Peptides-func benchmark \cite{LRGB}, AUPRC metric.}
\begin{tabular}{|c|cc|cc|cc|}
\hline
\multicolumn{1}{|c|}{\multirow{2}{*}{\textbf{Shuffle ratio}}} & \multicolumn{2}{c|}{\textbf{ECFP}}                      & \multicolumn{2}{c|}{\textbf{Amino acid counts}}          & \multicolumn{2}{c|}{\textbf{ESM2}}                       \\ \cline{2-7} 
\multicolumn{1}{|c|}{}                                      & \multicolumn{1}{l|}{\textbf{train}} & \textbf{train and test} & \multicolumn{1}{l|}{\textbf{train}} & \textbf{train and test} & \multicolumn{1}{l|}{\textbf{train}} & \textbf{train and test} \\ \hline
0.0                                                         & \multicolumn{1}{l|}{92.09}          & 92.09                   & \multicolumn{1}{l|}{91.02}          & 91.02                   & \multicolumn{1}{l|}{90.74}          & 90.74                   \\ \hline
0.1                                                         & \multicolumn{1}{l|}{91.73}          & 91.72                   & \multicolumn{1}{l|}{90.96}          & 91.04                   & \multicolumn{1}{l|}{90.40}           & 90.36                   \\ \hline
0.2                                                         & \multicolumn{1}{l|}{91.78}          & 91.94                   & \multicolumn{1}{l|}{89.89}          & 90.55                   & \multicolumn{1}{l|}{89.55}          & 90.52                   \\ \hline
0.3                                                         & \multicolumn{1}{l|}{91.03}          & 91.15                   & \multicolumn{1}{l|}{89.58}          & 90.47                   & \multicolumn{1}{l|}{88.69}          & 89.76                   \\ \hline
0.4                                                         & \multicolumn{1}{l|}{90.83}          & 91.24                   & \multicolumn{1}{l|}{87.81}          & 89.14                   & \multicolumn{1}{l|}{86.72}          & 89.74                   \\ \hline
0.5                                                         & \multicolumn{1}{l|}{90.61}          & 91.56                   & \multicolumn{1}{l|}{86.46}          & 88.76                   & \multicolumn{1}{l|}{86.83}          & 89.32                   \\ \hline
0.6                                                         & \multicolumn{1}{l|}{90.41}          & 91.91                   & \multicolumn{1}{l|}{85.99}          & 88.70                    & \multicolumn{1}{l|}{86.73}          & 89.26                   \\ \hline
0.7                                                         & \multicolumn{1}{l|}{89.73}          & 91.27                   & \multicolumn{1}{l|}{85.41}          & 89.03                   & \multicolumn{1}{l|}{85.04}          & 88.91                   \\ \hline
0.8                                                         & \multicolumn{1}{l|}{89.09}          & 90.94                   & \multicolumn{1}{l|}{83.71}          & 87.93                   & \multicolumn{1}{l|}{85.49}          & 89.44                   \\ \hline
0.9                                                         & \multicolumn{1}{l|}{88.84}          & 90.68                   & \multicolumn{1}{l|}{83.87}          & 88.57                   & \multicolumn{1}{l|}{85.28}          & 88.79                   \\ \hline
1.0                                                         & \multicolumn{1}{l|}{89.39}          & 90.91                   & \multicolumn{1}{l|}{83.83}          & 88.65                   & \multicolumn{1}{l|}{83.84}          & 89.50                    \\ \hline
\hline
delta                                                       & \multicolumn{1}{l|}{-2.7}           & -1.18                   & \multicolumn{1}{l|}{-7.19}          & -2.37                   & \multicolumn{1}{l|}{-6.9}           & -1.24                   \\ \hline
\end{tabular}
\label{lrgb_shuffling_results}
\end{table}

\begin{table}[h!]
\centering
\caption{Shuffling results for PeptideReactor benchmark \cite{PeptideReactor}, F1 metric.}
\begin{tabular}{|c|cc|cc|cc|}
\hline
\multicolumn{1}{|c|}{\multirow{2}{*}{\textbf{Shuffle ratio}}} & \multicolumn{2}{c|}{\textbf{ECFP}}                         & \multicolumn{2}{c|}{\textbf{Amino acid counts}}             & \multicolumn{2}{c|}{\textbf{ESM2}}                          \\ \cline{2-7} 
\multicolumn{1}{|c|}{}                                      & \multicolumn{1}{l|}{\textbf{train}} & \textbf{train and test} & \multicolumn{1}{l|}{\textbf{train}} & \textbf{train and test} & \multicolumn{1}{l|}{\textbf{train}} & \textbf{train and test} \\ \hline
0.0                                                         & \multicolumn{1}{l|}{79.39}          & 79.39                   & \multicolumn{1}{l|}{80.09}          & 80.09                   & \multicolumn{1}{l|}{76.88}          & 76.88                   \\ \hline
0.1                                                         & \multicolumn{1}{l|}{79.32}          & 79.43                   & \multicolumn{1}{l|}{78.80}           & 79.63                   & \multicolumn{1}{l|}{71.45}          & 75.81                   \\ \hline
0.2                                                         & \multicolumn{1}{l|}{78.74}          & 79.18                   & \multicolumn{1}{l|}{77.04}          & 78.57                   & \multicolumn{1}{l|}{63.66}          & 74.47                   \\ \hline
0.3                                                         & \multicolumn{1}{l|}{78.44}          & 79.18                   & \multicolumn{1}{l|}{76.80}          & 78.24                   & \multicolumn{1}{l|}{61.87}          & 74.11                   \\ \hline
0.4                                                         & \multicolumn{1}{l|}{77.72}          & 78.79                   & \multicolumn{1}{l|}{72.38}          & 77.01                   & \multicolumn{1}{l|}{62.69}          & 73.67                   \\ \hline
0.5                                                         & \multicolumn{1}{l|}{77.77}          & 79.21                   & \multicolumn{1}{l|}{70.39}          & 76.79                   & \multicolumn{1}{l|}{58.58}          & 73.32                   \\ \hline
0.6                                                         & \multicolumn{1}{l|}{77.44}          & 78.97                   & \multicolumn{1}{l|}{69.16}          & 76.57                   & \multicolumn{1}{l|}{54.49}          & 73.19                   \\ \hline
0.7                                                         & \multicolumn{1}{l|}{77.53}          & 79.02                   & \multicolumn{1}{l|}{68.46}          & 76.41                   & \multicolumn{1}{l|}{53.18}          & 73.19                   \\ \hline
0.8                                                         & \multicolumn{1}{l|}{77.36}          & 78.84                   & \multicolumn{1}{l|}{66.06}          & 76.38                   & \multicolumn{1}{l|}{58.29}          & 73.44                   \\ \hline
0.9                                                         & \multicolumn{1}{l|}{77.32}          & 78.81                   & \multicolumn{1}{l|}{65.28}          & 76.10                    & \multicolumn{1}{l|}{53.50}           & 73.04                   \\ \hline
1.0                                                         & \multicolumn{1}{l|}{77.60}          & 78.96                   & \multicolumn{1}{l|}{65.18}          & 76.11                   & \multicolumn{1}{l|}{59.38}          & 73.53                   \\ \hline
\hline
delta                                                       & \multicolumn{1}{l|}{-1.79}          & -0.43                   & \multicolumn{1}{l|}{-14.91}         & -3.98                   & \multicolumn{1}{l|}{-17.50}          & -3.35                   \\ \hline
\end{tabular}
\label{peptidereactor_shuffling_results}
\end{table}

\begin{table}[h!]
\centering
\caption{Shuffling results for XUAMP benchmark \cite{Xu_AMP}, AUROC metric.}
\begin{tabular}{|c|cc|cc|cc|}
\hline
\multicolumn{1}{|c|}{\multirow{2}{*}{\textbf{Shuffle ratio}}} & \multicolumn{2}{c|}{\textbf{ECFP}}                      & \multicolumn{2}{c|}{\textbf{Amino acid counts}}          & \multicolumn{2}{c|}{\textbf{ESM2}}                       \\ \cline{2-7} 
\multicolumn{1}{|c|}{}                                      & \multicolumn{1}{l|}{\textbf{train}} & \textbf{train and test} & \multicolumn{1}{l|}{\textbf{train}} & \textbf{train and test} & \multicolumn{1}{l|}{\textbf{train}} & \textbf{train and test} \\ \hline
0.0                                                         & \multicolumn{1}{l|}{75.01}          & 75.01                   & \multicolumn{1}{l|}{74.25}          & 74.25                   & \multicolumn{1}{l|}{77.10}           & 77.10                    \\ \hline
0.1                                                         & \multicolumn{1}{l|}{75.1}           & 75.24                   & \multicolumn{1}{l|}{74.2}           & 74.07                   & \multicolumn{1}{l|}{76.41}          & 76.16                   \\ \hline
0.2                                                         & \multicolumn{1}{l|}{74.35}          & 74.35                   & \multicolumn{1}{l|}{72.94}          & 72.63                   & \multicolumn{1}{l|}{75.30}           & 74.59                   \\ \hline
0.3                                                         & \multicolumn{1}{l|}{73.06}          & 73.34                   & \multicolumn{1}{l|}{73.38}          & 73.45                   & \multicolumn{1}{l|}{74.65}          & 72.74                   \\ \hline
0.4                                                         & \multicolumn{1}{l|}{71.66}          & 73.57                   & \multicolumn{1}{l|}{72.25}          & 72.07                   & \multicolumn{1}{l|}{74.91}          & 72.34                   \\ \hline
0.5                                                         & \multicolumn{1}{l|}{71.66}          & 73.29                   & \multicolumn{1}{l|}{72.89}          & 72.45                   & \multicolumn{1}{l|}{73.33}          & 71.44                   \\ \hline
0.6                                                         & \multicolumn{1}{l|}{72.41}          & 73.13                   & \multicolumn{1}{l|}{72.46}          & 72.21                   & \multicolumn{1}{l|}{72.92}          & 70.76                   \\ \hline
0.7                                                         & \multicolumn{1}{l|}{72.55}          & 72.9                    & \multicolumn{1}{l|}{72.10}           & 72.69                   & \multicolumn{1}{l|}{72.84}          & 70.09                   \\ \hline
0.8                                                         & \multicolumn{1}{l|}{72.78}          & 73.25                   & \multicolumn{1}{l|}{72.15}          & 72.51                   & \multicolumn{1}{l|}{69.86}          & 70.48                   \\ \hline
0.9                                                         & \multicolumn{1}{l|}{72.81}          & 73.38                   & \multicolumn{1}{l|}{72.31}          & 71.74                   & \multicolumn{1}{l|}{71.06}          & 69.86                   \\ \hline
1.0                                                         & \multicolumn{1}{l|}{73.41}          & 73.35                   & \multicolumn{1}{l|}{71.63}          & 71.67                   & \multicolumn{1}{l|}{71.25}          & 70.61                   \\ \hline
\hline
delta                                                       & \multicolumn{1}{l|}{-1.60}           & -1.66                   & \multicolumn{1}{l|}{-2.62}          & -2.58                   & \multicolumn{1}{l|}{-5.85}          & -6.49                   \\ \hline
\end{tabular}
\label{xuamp_shuffling_results}
\end{table}

\clearpage

\section{Analysis of difficult predictions and errors}
\label{appendix_difficult_predictions}

To investigate whether certain data samples are systematically more difficult for the model to predict, we used the AutoPeptideML benchmark \cite{AutoPeptideML}. This benchmark provides multiple prediction tasks with independent datasets and challenging train-test splits.

\begin{figure}[t]
    \centering
    \includegraphics[width=\textwidth]{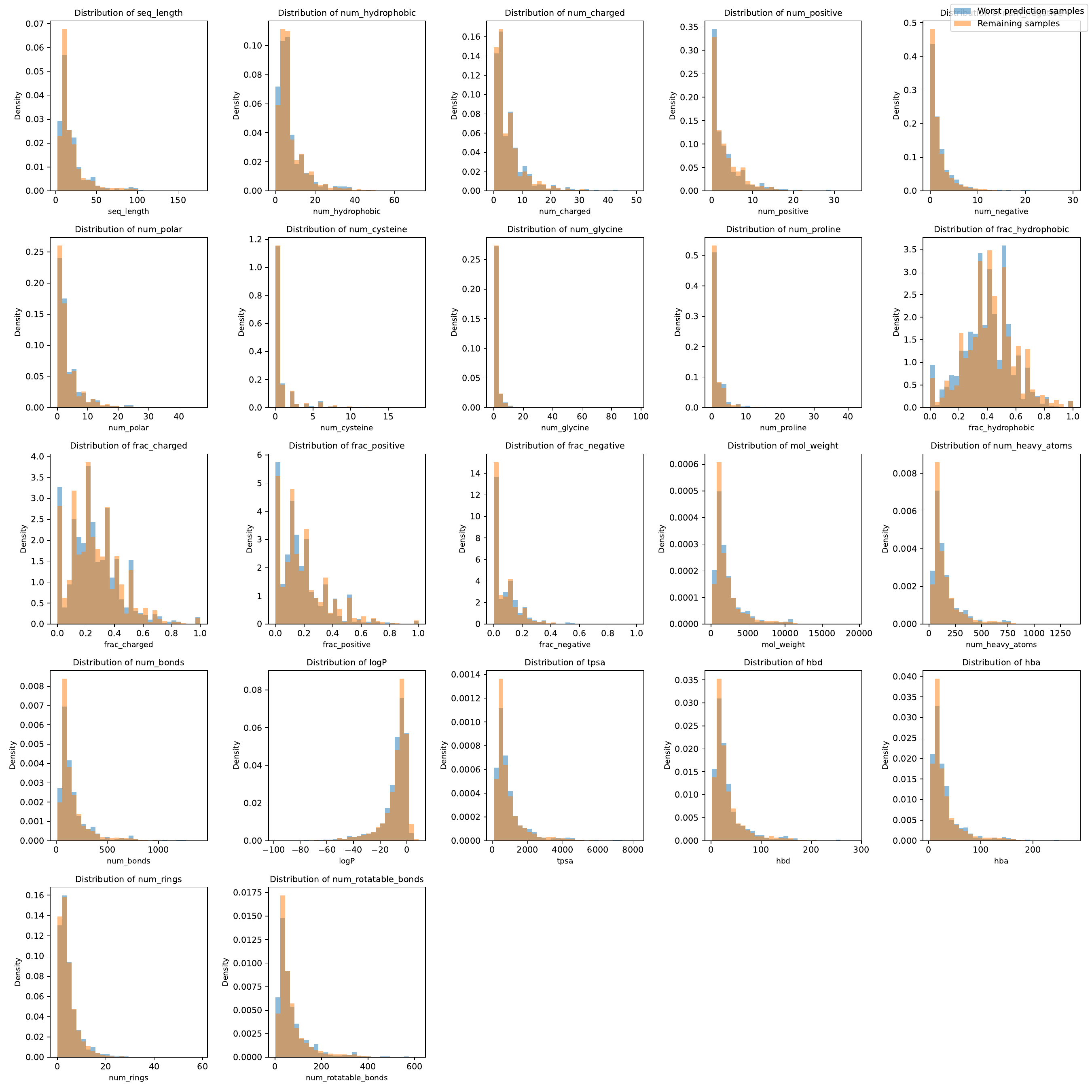}
    \caption{AutoPeptideML \cite{AutoPeptideML} feature distributions.}
    \label{fig:autopeptideml_feature_distributions}
\end{figure}

For each dataset, we selected samples for which the absolute difference between the true class label and the predicted probability exceeded 0.9. We compared the distributions of their physicochemical properties with those of the remaining samples, with smaller error, summarized in Figure \ref{fig:autopeptideml_feature_distributions}. Descriptors included: sequence length; counts and fractions of positively charged, negatively charged, charged, and polar amino acids, counts of cysteines, glycines, and prolines. Additionally, we use more features typical in chemoinformatics, such as molecular weight, number of heavy atoms and bonds, lipophilicity, topological polar surface area, numbers of hydrogen bond donors and acceptors, numbers of rings and number of rotatable bonds.

We use the Epps-Singleton statistical test \cite{epps_singleton_test} to check if there is a statistically significant difference between the two distributions for each descriptor. It was selected as we have both continuous and discrete descriptors, and this test has higher power for discrete ones than Kolmogorov-Smirnov. Because the number of high-error samples is much smaller than the remaining samples, we repeated the test 100 times using the full high-error set and bootstrap samples of the remaining data matched in size.

Out of 16 AutoPeptideML datasets (tasks), only 3 showed statistically significant differences in any feature, and no feature was consistently significant across tasks. We additionally aggregated prediction errors across the entire benchmark and repeated the analysis, which again revealed no statistically meaningful differences between the distributions of the worst predictions and the remaining samples.

Overall, these results indicate that no individual physicochemical property consistently increases prediction difficulty when using fingerprint-based representations.

\clearpage

\section{Sequence length split}

To evaluate robustness of models to length of training and testing data, we use sequence length split. We selected 3 datasets for this experiment: AutoPeptideML, Peptides-func from LRGB, and XUAMP, and we test 3 models: ECFP, amino acid counts, and ESM2.

Peptides in each dataset are sorted by sequence length and divided into 5 length buckets: 20\% of the shortest ones, those of lengths 20-40\% of distribution, and so on. Each one has the same number of peptides, and we perform a cross-validation on those lengths. For example, when we test on 0-20\% length sequences (i.e., 20\% of shortest ones), we use all others for training. As this kind of splitting can introduce heavy class imbalance, we use MCC metric as well suited for imbalanced classification \cite{MCC}, and also report the average positive class percentage (averaged over datasets for AutoPeptideML).

Results are summarized in Tables \ref{table_seq_length_split_autopeptideml}, \ref{table_seq_length_split_lrgb}, and \ref{table_seq_length_split_xuamp}. Overall, all models exhibit quite similar behavior, with the highest results in the middle 40-60\% test bucket, where they can interpolate more from the training data. Results are universally lower for shortest peptides, which contain the least information that the model can use. The performance of molecular fingerprints indicates that they are not particularly negatively affected by peptide size, performing on par with sequence-based baselines.

\begin{table}[th!]
\centering
\caption{Sequence length split results for AutoPeptideML benchmark.}
\begin{tabular}{|c|c|c|c|c|c|}
\hline
\textbf{Model} & \textbf{0-20\%} & \textbf{20-40\%} & \textbf{40-60\%} & \textbf{60-80\%} & \textbf{80-100\%} \\ \hline
ECFP & 0.260 & 0.458 & 0.482 & 0.489 & 0.376 \\ \hline
Amino acid counts & 0.290 & 0.409 & 0.463 & 0.400 & 0.355 \\ \hline
ESM2 & 0.262 & 0.435 & 0.508 & 0.527 & 0.385 \\ \hline
\hline
\begin{tabular}[c]{@{}c@{}}Average positive\\ class percentage\end{tabular} & 52\% & 51\% & 47\% & 51\% & 48\% \\ \hline
\end{tabular}
\label{table_seq_length_split_autopeptideml}
\end{table}

\begin{table}[th!]
\centering
\caption{Sequence length split results for LRGB Peptides-func benchmark.}
\begin{tabular}{|c|c|c|c|c|c|}
\hline
\textbf{Model} & \textbf{0-20\%} & \textbf{20-40\%} & \textbf{40-60\%} & \textbf{60-80\%} & \textbf{80-100\%} \\ \hline
ECFP & 0.149 & 0.368 & 0.418 & 0.321 & 0.215 \\ \hline
Amino acid counts & 0.230 & 0.419 & 0.452 & 0.393 & 0.268 \\ \hline
ESM2 & 0.099 & 0.337 & 0.377 & 0.293 & 0.240 \\ \hline
\hline
\begin{tabular}[c]{@{}c@{}}Average positive\\ class percentage\end{tabular} & 13\% & 17\% & 18\% & 17\% & 17\% \\ \hline
\end{tabular}
\label{table_seq_length_split_lrgb}
\end{table}

\begin{table}[th!]
\centering
\caption{Sequence length split results for XUAMP benchmark.}
\begin{tabular}{|c|c|c|c|c|c|}
\hline
\textbf{Model} & \textbf{0-20\%} & \textbf{20-40\%} & \textbf{40-60\%} & \textbf{60-80\%} & \textbf{80-100\%} \\ \hline
ECFP & 0.427 & 0.577 & 0.540 & 0.483 & 0.427 \\ \hline
Amino acid counts & 0.285 & 0.525 & 0.532 & 0.496 & 0.409 \\ \hline
ESM2 & 0.285 & 0.648 & 0.675 & 0.601 & 0.470 \\ \hline
\hline
\begin{tabular}[c]{@{}c@{}}Average positive\\ class percentage\end{tabular} & 96\% & 57\% & 42\% & 31\% & 22\% \\ \hline
\end{tabular}
\label{table_seq_length_split_xuamp}
\end{table}
\clearpage

\section{Sequence motif detection}

There are tasks that are short-range on sequence level, yet long-range on the molecular graph level. They are challenging for the proposed molecular fingerprints approach, and we provide an example of such as task as an example of peptide therapeutic design task that benefits more from more common baselines. This delineates the applicability domain of the proposed model - if the task requires incorporating order of amino acids, particularly those far from each other in the molecular graph, the sequence-based models have a strong advantage.

The task is recognizing highly charged sequence motifs ``KKK'' and ``RRR'' in sequence, which are known to impact peptide-protein binding and are highly relevant to peptide therapeutic design \cite{sequence_motif_KKK,sequence_motif_KKK_2,sequence_motif_RRR}. This is basically checking if there is a substring in the amino acid sequence, and thus is trivial for PLMs like ESM2. It is much harder, but not impossible, for molecular fingerprints, as they ignore order of subgraphs, and thus have very weak learning signal in this task.

We use XUAMP dataset, with the same train and test splits, but replace the original labels with 1 if ``KKK'' or ``RRR'' is present in sequence, or 0 otherwise. About 8\% of data has positive labels, so we report MCC metric, which works well for imbalanced cases \cite{MCC}. Molecular fingerprints and amino acid counts use the same models as in the main body, while ESM2-35M is finetuned, in order to show its maximal possible performance. Our goal here is not to achieve any particular performance, but rather show the ability of those methods to incorporate long-range dependencies. Finetuning is kept lightweight and short, with just 2 epochs and learning rate $5*10^{-5}$, optimizing BCE loss. 

Results are summarized in Table \ref{table_sequence_motifs}. As expected, molecular fingerprints achieve results better than random, but not great. Amino acid counts baseline performs well, while ESM2 gets a near-perfect result.

\begin{table}[ht!]
\centering
\caption{Results on sequence motif detection task.}
\begin{tabular}{|c|c|}
\hline
\textbf{Model} & \textbf{MCC} \\ \hline
ECFP & 0.270 \\ \hline
Topological Torsion & 0.267 \\ \hline
RDKit & 0.238 \\ \hline
Amino acid counts & 0.616 \\ \hline
ESM2 & 0.993 \\ \hline
\end{tabular}
\label{table_sequence_motifs}
\end{table}

\end{document}